\documentclass[aps,prb,showpacs,twoside,twocolumn,10pt]{revtex4-2}
\usepackage[colorlinks=true, citecolor=blue, urlcolor=blue]{hyperref}
\usepackage{times,epsfig,amssymb,amsfonts,amsmath,bm,subfigure,mathtools,amsthm,braket,soul,enumitem,xcolor,physics,graphics,graphicx,tikz}
\UseRawInputEncoding
\usepackage[normalem]{ulem}
\usepackage{comment}
\usepackage{epstopdf}
\usepackage{relsize}
\usepackage{orcidlink}
\newtheorem{theorem}{Theorem}

\newtheorem{proposition}[theorem]{Proposition}

\begin{document}

\title{Interplay of Nonstabilizerness and Ergotropy in Quantum Batteries}
 
\author{Tanoy Kanti Konar$^{1}$\orcidlink{0000-0002-8188-0745}}
\email{tanoy.kanti.konar@uj.edu.pl}
\author{Jakub Zakrzewski$^{1,2}$\orcidlink{0000-0003-0998-9460}}
\email{jakub.zakrzewski@uj.edu.pl}

\affiliation{$^{1}$Instytut Fizyki Teoretycznej, Wydzia\l{} Fizyki, Astronomii i Informatyki Stosowanej, Uniwersytet Jagiello\'nski, \L{}ojasiewicza 11, PL-30-348 Krak\'ow, Poland}
\affiliation{$^{2}$Mark Kac Complex Systems Research Center, Jagiellonian University in Krak\'ow, PL-30-348 Krak\'ow, Poland}

\begin{abstract}
Nonstabilizerness plays an essential role in the efficient simulation of quantum systems on quantum computers. In this work, we investigate its role in the context of quantum batteries (QBs). To this end, we consider a system of $N$ spin-$\tfrac{1}{2}$ particles, with the left half serving as the charger and the right half as the battery. By studying different classes of interactions between the charger and the battery, we quantify the amount of nonstabilizerness required to store work in the QB. Our results reveal a strong correlation between the ergotropy stored in the battery and the total nonstabilizerness of the composite system whenever the Hamiltonian preserves a $\mathbb{U}(1)$ symmetry. In contrast, this correspondence is generally lost for more generic interactions. A similar behavior is observed for unitary circuit dynamics. Finally, we examine the complementary scenario in which the battery is initialized in a nonstabilizer state and subsequently charged through Clifford evolution. In this case, we find that the maximum average charging power is nonmonotonic in the initial nonstabilizerness. Remarkably, the highest average power can be achieved even when the initial state carries no magic (nonstabilizerness), demonstrating that initial magic is not a necessary ingredient for generating optimal charging power.
\end{abstract}

\maketitle

\section{Introduction}
\label{sec:intro}
Quantum batteries (QBs) have emerged as a promising paradigm in modern quantum technologies, functioning as energy storage devices governed by the principles of quantum mechanics~\cite{Alicki2013,quantum_battery_review}. Since their inception, substantial progress has been made in uncovering mechanisms that enable quantum advantage in such systems, as well as in exploring their realization across a wide range of physical platforms~\cite{Binder2015,Campaioli2017,Ferraro2018,Andolina_prl_2019,caravelli_prr_2020,Shi2022,Konar2022,Farre2020,Yang2023,imai_pra_2023,Bhanja_pra_2024,vigneshwar2025}. In particular, diverse architectures have been proposed in settings such as quantum spin chains~\cite{Le_pra_2018,rossini_prb_2019,Ghosh2020,mondal_pre_2022,guo_pra_2024,Zhang2024,Gyhm2024Jan,puri2025,sahoo2025,Bhattacharya2026} and atom-cavity systems~\cite{lu_pra_2021,dou_pra_2022,zhao_pra_2025,pushpan2025}. Moreover, it has been shown that environmental noise can play a dual role in charging dynamics, either enhancing or degrading the performance of a QB depending on the specific physical scenario~\cite{Ghosh2021,Zakavati2021,Arjmandi2022,Liu2024,Tirone2023,sarkar2025,Kamin2020,Santos2021,Xu2024,Chaki2025,Morrone2023,ahmadi_prl_2024,Medina2024,Maryam2025,topological_quantumbattery,Vigneshwar2026}. Beyond conceptual proposals, several experimental platforms have demonstrated proof-of-principle realizations of quantum batteries, including nuclear magnetic resonance (NMR)~\cite{MaheshexpNMR}, quantum dots~\cite{wennigerexpqdots}, organic semiconductors~\cite{recentexperiment}, and superconducting circuits~\cite{superconducting_battrey_1,superconductQBexp,GemmeexpIBMsupercond}. Also, proposals have been put forward to perform quantum computation using QBs~\cite{kurman_prx_2026}. It is now well understood that the physical resource responsible for quantum advantage in quantum batteries, such as entanglement or coherence, is not universal~\cite{gyhm_prl_2022}, but instead depends sensitively on the underlying charging Hamiltonian.

\begin{figure}
    \centering
    \includegraphics[width=\linewidth]{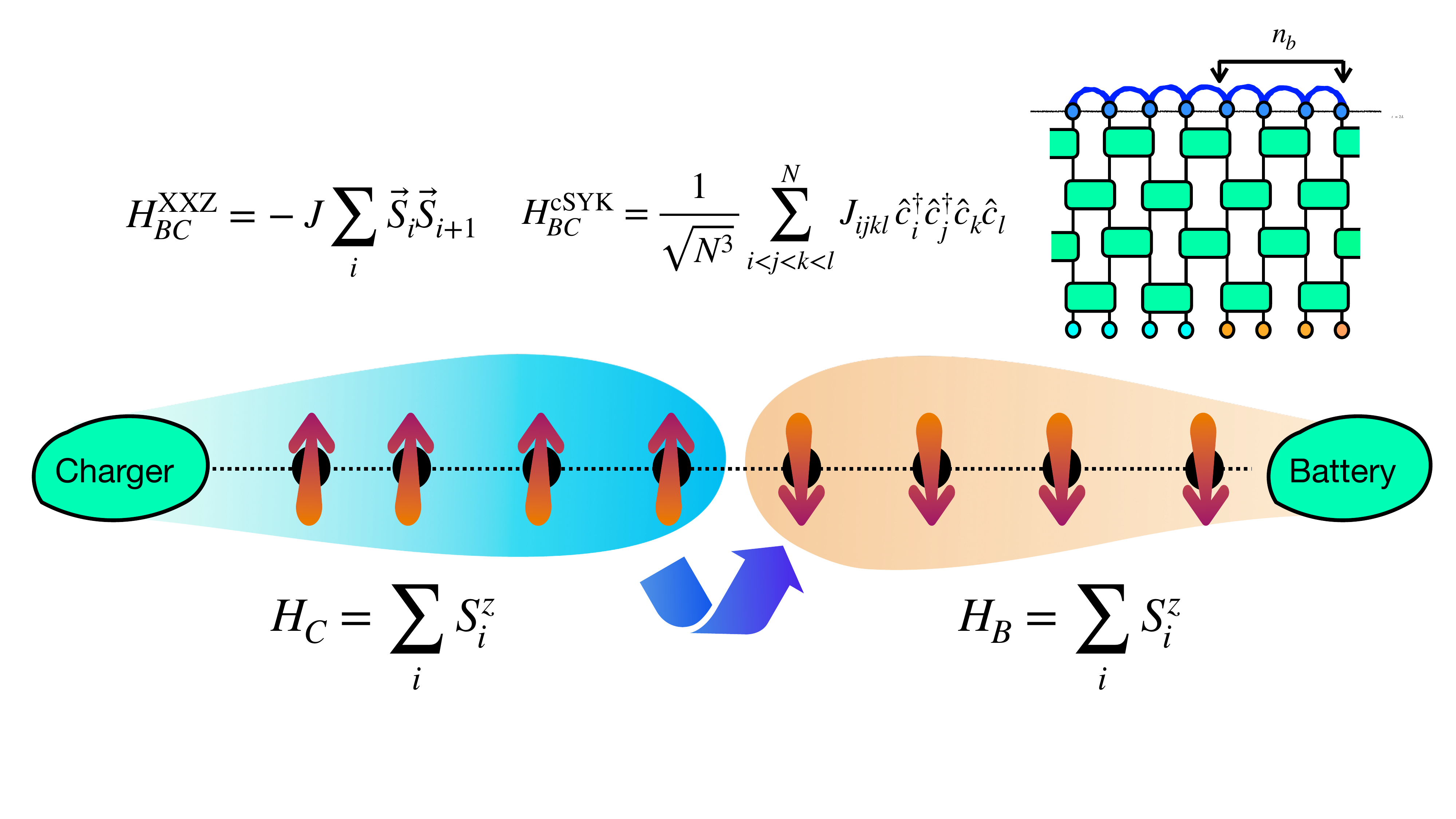}
    \caption{\textbf{Schematic illustration of the quantum battery (QB) and charger setup.} We consider a one-dimensional chain of spin-$\tfrac{1}{2}$ particles comprising both the charger and the battery. The left half of the chain serves as the charger, while the right half functions as the quantum battery. The initial state of the composite system is prepared as \(\ket{\psi(0)} = \ket{\uparrow}^{\otimes N/2} \ket{\downarrow}^{\otimes N/2}\), such that the charger is fully excited and the battery is in its ground state. The total system is then evolved under different interaction Hamiltonians that mediate energy transfer between the two subsystems. Our goal is to quantify the amount of nonstabilizerness generated in the full system in order to store an amount of work in the quantum battery.}
    \label{fig:schematics}
\end{figure}

Quantum entanglement~\cite{horodecki_review}, although widely regarded as a genuinely quantum property, is by itself not sufficient to enable universal quantum computation. For example, many stabilizer states can exhibit a high degree of entanglement while remaining efficiently simulable on a classical computer~\cite{aaronson_pra_2004,nielsen2010}. In this context, quantum nonstabilizerness, commonly referred to as ``magic,'' emerges as an essential resource for characterizing computational complexity~\cite{bravyi_pra_2005}. It quantifies the extent to which a quantum protocol departs from polynomial-time classical simulability within the framework of the Gottesman--Knill theorem~\cite{gottesman1998}. Several protocols have been proposed that demonstrate genuine quantum advantage enabled by the presence of nonstabilizerness~\cite{gottesman1997,gottesman_pra_1998,aaronson_pra_2004}. However, accurately quantifying the amount of nonstabilizerness in a quantum state is often computationally demanding. A prominent example is the robustness of magic (RoM)~\cite{howard_prl_2017}, whose evaluation becomes prohibitively expensive for large system sizes. Recently, significant progress has been achieved through the introduction of stabilizer R\'enyi entropies (SRE)~\cite{leone_prl_2022}, which provide a computationally tractable measure of nonstabilizerness, particularly for pure states. In the many-body setting, several works have further suggested an intriguing connection between nonstabilizerness and quantum criticality~\cite{white_prb_2021,Sarkar2020,oliviero_pra_2022,haug_prb_2023,tarabunga_prxq_2023,Tarabunga2024}, Kibble--Zurek scaling~\cite{grabarits2026}, and many-body localization~\cite{falcao_prl_2025,li2026}. From the perspective of quantum technology, nonstabilizerness has been extensively explored in the contexts of quantum computation~\cite{Campbell2017,Bravyi2019,Oliviero2022}, quantum algorithms~\cite{capecci2025}, and quantum metrology~\cite{yanes_pra_2026}, further underscoring its fundamental importance as a key operational resource in emerging quantum devices.

In this work, we investigate the role of nonstabilizerness in storing both energy and ergotropy in QBs. From the perspective of resource theory, the ergotropy and nonstabilizerness correspond to distinct resources, and in general no direct relation exists between them. For example, a battery can be charged entirely through Clifford unitaries without generating additional magic. However, realistic charging protocols are usually not restricted to Clifford operations, which can produce magic.  To address this issue, we consider a one-dimensional chain of $N$ spin-$\tfrac{1}{2}$ particles, where the left half acts as the charger and the right half serves as the battery. The charger is initially prepared in the fully excited state, while the battery is initialized in its ground state. The system then evolves under a global charging Hamiltonian, transferring energy from the charger to the battery. Our goal is to establish the relation between the stabilizer R\'enyi entropy (SRE), which quantifies nonstabilizerness, and the work and ergotropy stored in the battery. We first study two interaction models: the \(\mathrm{XXZ}\) spin chain and the complex Sachdev-Ye-Kitaev (\(\mathrm{cSYK}\)) model. 
Further, we also consider a brick-wall circuit protocol where the charging dynamics is generated by different classes of two-qubit unitaries, including generic Haar-random unitaries, $\mathbb{U}(1)$-symmetric Haar-random unitaries, and Hamiltonian-generated gates.
￼

The paper is organized as follows. In Sec.~\ref{sec:battery_magic}, we introduce nonstabilizerness and describe the charging protocol of the quantum battery. In Secs.~\ref{sec:xxz_battery} and \ref{sec:syk_battery}, we present a detailed analysis of the relation between the stabilizer R\'enyi entropy (SRE) and ergotropy for the \(\mathrm{XXZ}\) and \(\mathrm{cSYK}\) interaction models, respectively. In Sec.~\ref{sec:magic_brick_wall}, we extend our study beyond specific interaction Hamiltonians by considering a brick-wall circuit architecture with different classes of two-qubit unitaries. In Sec.~\ref{sec:initial_magic}, we investigate the role of initial nonstabilizerness in determining the maximum average charging power of the battery. Finally, in Sec.~\ref{sec:conclusion}, we summarize our main results and conclude the paper.

\section{Formulation of the problem}
\label{sec:battery_magic}
Let us first define the general scheme of the quantum battery as well as the observables that will be addressed in the following.

\subsection{The battery}
To investigate the behavior of magic during the charging process, we consider a composite setup consisting of a battery and a charger. The battery Hamiltonian is given by
\begin{equation}
    H_B = \sum_{i=1}^{n_b} S_z^i,
    \label{eq:battery_hamiltonian}
\end{equation}
where $S_z^i = \sigma_z^i/2$, with $\sigma_z^i$ being the \(z\)-component of the Pauli matrix acting on site $i$, and $n_b$ denotes the number of spins in the battery. The battery is initially prepared in its ground state, which is a fully polarized product state,
\begin{equation}
    \ket{\psi_B} = \ket{\downarrow \dots \downarrow},
    \label{eq:initial_state}
\end{equation}
where $\sigma_z \ket{\downarrow} = - \ket{\downarrow}$. This state is completely uncorrelated and has vanishing magic.

The charger is modeled as an identical spin system of size $n_b$, with Hamiltonian \(H_C = \sum_{i=1}^{n_b} S_z^i\), and is initially prepared in the fully excited state, \(\ket{\psi_C} = \ket{\uparrow \dots \uparrow}\), where $\sigma_z \ket{\uparrow} = + \ket{\uparrow}$. Together, the charger and battery form a composite system of $N = 2n_b$ spins, where the left half acts as the charger and the right half serves as the battery. The initial state of the total system is thus given by 
\begin{equation}
\ket{\psi(0)}=\ket{\psi_C}\otimes \ket{\psi_B}. 
\label{inistate}
\end{equation}
The interaction between these two subsystems enables energy transfer, thereby charging the QB (see Fig.~\ref{fig:schematics}).

\noindent\emph{Interaction between battery and charger.}
After preparing the initial state, an interaction between the charger and the battery is switched on, governed by the Hamiltonian $H_{BC}$. The total charging Hamiltonian of the composite system is therefore given by
\begin{equation}
    H_t = H_B + H_C + H_{BC}.
\end{equation}
In this work, we consider different classes of interaction Hamiltonians, such as the \(\mathrm{XXZ}\) model, the \(\mathrm{cSYK}\) model, and random unitary dynamics. These interactions facilitate the transfer of both energy and ergotropy from the charger to the battery. Our primary objective is to find possible relations between the amount of magic created in the process of charging the battery and typical observables used in QB efficiency descriptions, such as the energy stored in the battery, $W$, and the ergotropy, $\mathcal{E}$, in the QB.

\subsection{Nonstabilizerness} 
Nonstabilizerness (quantum magic) quantifies the complexity of preparing a given quantum state, or equivalently, the difficulty of simulating it on a classical computer. Here, we shall be interested in the creation of quantum magic during the process of charging the QB. Several measures of nonstabilizerness have been introduced in the literature, including the robustness of magic (RoM)~\cite{howard_prl_2017} and mana~\cite{Veitch2014}. In this work, we focus on the stabilizer R\'enyi entropy (SRE) as a quantifier of nonstabilizerness for the full system, comprising both the charger and the battery. This choice is purely pragmatic: for the many-body spin chains discussed, measures other than SRE are much more difficult to calculate.

For a pure state $\ket{\psi}$, the SRE is defined as
\begin{equation}
    \mathcal{M}_\alpha(\ket{\psi}) 
    = \frac{1}{1-\alpha} \log_2 \left[
    \frac{1}{2^N} \sum_{P \in \mathcal{P}_N} 
    \langle \psi | P | \psi \rangle^{2\alpha}
    \right],
\end{equation}
where $\mathcal{P}_N$ denotes the $N$-qubit Pauli group and $N$ is the total number of spins in the system. Although the SRE is relatively straightforward to compute compared to other measures of nonstabilizerness, numerical limitations become significant for larger system sizes due to the exponential growth of the Hilbert space. In this work, we employ the \emph{HadaMAG.jl} package for the efficient calculation of SRE~\cite{Sierant2026} and consider only the $\alpha=2$ case, i.e., $\mathcal{M}_2$.

\subsection{Work and ergotropy} 
The amount of energy stored in the quantum battery is quantified by
\begin{equation}
    W(t)=\Tr[H_B\rho(t)]-\Tr[H_B\rho(0)],
\end{equation}
where \(\rho(0)\) is the initial state of the battery, \(\rho(t)\) is the time-evolved state, and \(\Tr[H_B\rho(t)]\) represents the energy of the battery at time \(t\). For closed systems, this energy is equivalent to the work stored in the battery.
The maximum amount of extractable work through unitary operations, known as \emph{ergotropy}~\cite{Alicki2013}, is defined as
\begin{equation}
\mathcal{E}(\rho(t)) = \Tr[\rho(t) H_B] - \min_{U}\Tr[U\rho(t)U^{\dagger}H_B],
\label{eq:ergotropy_def}
\end{equation}
where the minimization is performed over all possible unitary operators \(U\). The second term in Eq.~(\ref{eq:ergotropy_def}) corresponds to the energy of the associated passive state \(\rho_{\pi}\), i.e., the state from which no further work can be extracted by any unitary transformation~\cite{Allahverdyan2004}.
Assume that the battery Hamiltonian is written as
\(H_B=\sum_{i}\epsilon_i |\epsilon_i\rangle \langle \epsilon_i|\), with the eigenenergies ordered as \(\epsilon_i\geq \epsilon_{i-1}\) for all \(i\).
For a density matrix written in its spectral decomposition as \(\rho=\sum_{i=1}^{n} p_i |p_i\rangle \langle p_i|\),
with eigenvalues ordered as \(p_i\geq p_{i+1}\) for all \(i\), the passive state is given by \(\rho_{\pi} = \sum_{i=1}^{n} p_i |\epsilon_i\rangle \langle \epsilon_i|\), where \(\{|\epsilon_i\rangle\}\) are the eigenstates of the battery Hamiltonian. In this construction, the largest occupation probabilities are assigned to the lowest-energy states, ensuring that \(\rho_{\pi}\) is passive. Using this, Eq.~(\ref{eq:ergotropy_def}) reduces to \(\mathcal{E}(\rho) = \Tr[\rho(t) H_B]-\sum p_i \epsilon_i\).

\noindent\emph{Average SRE and ergotropy.} 
In QBs, it is possible that the maximally energetic state is itself a stabilizer state. However, even when both the initial ground state and the final fully charged state are stabilizer states, the intermediate dynamics can generate nonstabilizerness. Therefore, it is important to analyze not only the instantaneous behavior of nonstabilizerness and ergotropy, but also their time-averaged values during the charging process. To capture this behavior, we introduce the time-averaged quantity as a relevant figure of merit, defined as
\begin{equation}
    \langle X \rangle = \frac{1}{t} \int_0^t X(t') \, \mathrm{d}t',
    \label{eq:avg_figs_of_merit}
\end{equation}
where $X \in \{\mathcal{M}_2, \mathcal{E}, W\}$ corresponds to the second-order SRE, ergotropy, and stored energy, respectively. Furthermore, to quantify the total energy that can be stored in the quantum battery, we also consider the asymptotic long-time limit $t \to \infty$, which characterizes the steady-state or saturation value of the stored energy.

\section{Nonstabilizerness growth and energy storage via \(\mathrm{XXZ}\) spin chain}
\label{sec:xxz_battery}
In this section, we investigate the relationship between nonstabilizerness and the amount of energy stored in a quantum battery when the battery and charger interact via an \(\mathrm{XXZ}\) spin chain. After preparing the initial state, we switch on interactions between the spins and consider two distinct types of interaction Hamiltonians. We first focus on a nearest-neighbor interaction described by the spin-$\tfrac{1}{2}$ \(\mathrm{XXZ}\) model, given by
\begin{equation}
    H_{BC}^{\mathrm{XXZ}} = -J \sum_{i=1}^{N-1} \vec{S}^{\,i} \cdot \vec{S}^{\,i+1},
    \label{eqn:total_h}
\end{equation}
where $\vec{S}^{\,i} \in \{\sigma_x^i/2,\, \sigma_y^i/2,\, \sigma_z^i/2\}$ are the spin operators. This Hamiltonian possesses a $\mathbb{U}(1)$ symmetry associated with the conservation of total magnetization. As a result, the total energy of the system is conserved, and any energy stored in the battery arises exclusively from energy transferred from the charger (cf. Ref.~\cite{tirrito2025}).

The initial state, Eq. \eqref{inistate}, represents a single domain wall in the spin chain. Due to the local character of interactions for the \(\mathrm{XXZ}\) Hamiltonian, only spins at the domain wall are initially affected by the dynamics; the remaining spins remain essentially frozen. Therefore, the initial dynamics is captured by the two-spin model. Let us, therefore, consider such a model first.

\subsection{Two-qubit system}
In this minimal system, a single spin acts as a charger and the second one as a battery. The initial state is given as \(\ket{\Psi(0)}=\ket{\uparrow\downarrow}\), and it evolves under the Hamiltonian in Eq. \ref{eqn:total_h}, which can effectively be written as 
\begin{equation}
    H_{\mathrm{eff}} =J
    \begin{bmatrix}
        \frac{1}{4} & -\frac{1}{2} \\
        -\frac{1}{2} & \frac{1}{4}
    \end{bmatrix}.
\end{equation}
The diagonal contribution arising from the $ZZ$-interaction merely produces a global phase in the evolution and can be removed. The time-evolved state takes the form
\begin{equation}
    \ket{\psi(t)} = a(t)\ket{\uparrow\downarrow} + b(t)\ket{\downarrow\uparrow},
    \label{eq:effective_evolution}
\end{equation}
and the coefficients are given by
\begin{equation}
    a(t) = \cos\left(\frac{Jt}{2}\right), 
    \qquad 
    b(t) = -i \sin\left(\frac{Jt}{2}\right).
\end{equation}
The probability that the spin flips from $\ket{\downarrow}$ to $\ket{\uparrow}$ is therefore
\begin{equation}
    p(t) = |b(t)|^2 = \sin^2\left(\frac{Jt}{2}\right)
    \label{mwork}
\end{equation}
and gives the energy stored in the battery, given as  $W=p(t)$. 
\begin{figure}
    \centering
    \includegraphics[width=\linewidth]{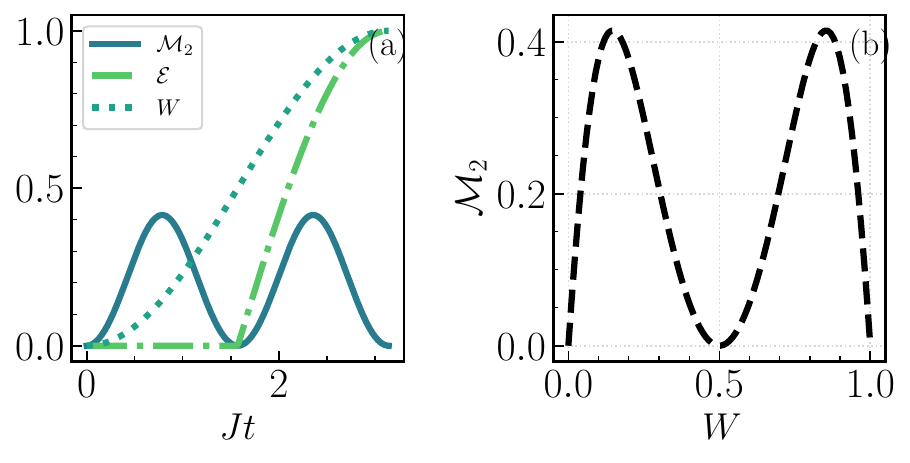}
    \caption{{\bf The stored work, $W$; the ergotropy, $\mathcal{E}$, and the SRE, $\mathcal{M}_2$ in the two-qubit system.} Panel (a): Time dependence of these observables are presented; (b): SRE versus work, $W$. All axes are dimensionless.}
    \label{fig:magic_work}
\end{figure}
In a similar spirit, we find the time dependence of SRE. The sum of the expectation values to the fourth power over all elements of the Pauli group is given by
\begin{equation}
\sum_{P}|\langle P\rangle|^4=4-16p(1-p)(1-2p)^2.
\end{equation}
Now, since $p(t)=W$, the SRE of the state is given by
\begin{equation}
    \mathcal{M}_2(W)=-\log_2\big [ 1-4W(1-W)(1-2W)^2\big ],
    \label{eq:magic_work_conn}
\end{equation}
identifying the exact relation between the SRE generated in the process and the work stored in the battery. The explicit time dependence reads
\begin{equation}
    \mathcal{M}_2(t)=-\log_2\big [1-\frac{1}{4} \sin^2(2Jt)].
\end{equation}
Similarly, we find the time dependence of the ergotropy. The battery (the second spin) is diagonal in the energy basis. As long as $p(t)<0.5$, the battery remains passive and its ergotropy is zero. The ergotropy is given analytically as
\begin{equation}
    \mathcal{E}(t)= \frac{1}{2}\big [\abs{\cos Jt}-\cos Jt \big] .
\end{equation}
Note that the ergotropy may also be expressed as a function of $W$:
\begin{equation}
    \mathcal{E}(t)= \frac{1}{2}\big [\abs{1-2W}-(1-2W) \big] .
\end{equation}
Both $W$ and $\mathcal{E}$ undergo Bloch oscillations with a circular frequency $J$, while the SRE oscillates twice as fast -- compare Fig.~\ref{fig:magic_work}(a). This is due to the fact that nonstabilizerness is maximal for states that are an equal superposition of the $\ket{\uparrow\downarrow}$ and $\ket{\downarrow\uparrow}$ states, which occurs twice during a single Bloch oscillation of the two-level system. 

This minimal model demonstrates that no simple monotonic relationship connects ergotropy and nonstabilizerness. On the other hand, a direct quantitative relation emerges between the work stored in the battery and the amount of global nonstabilizerness generated during the charging process (see Eq.~(\ref{eq:magic_work_conn})), revealing that storing an amount $W$ of energy requires $\mathcal{M}_2(W)$ of SRE. As illustrated in Fig.~\ref{fig:magic_work}(b), even when the battery reaches its maximum stored work ($W = 1$), the global nonstabilizerness vanishes entirely. Interestingly, the point of maximum nonstabilizerness does not coincide with the point of maximum stored work, showing that optimal charging does not require maximal magic generation. In Fig.~\ref{fig:magic_work}, the dynamics is shown up to $Jt = \pi$, corresponding to a full charging of the battery, after which the roles of the battery and the charger reverse periodically.

\subsection{XXZ model for \(N> 2\)}

\begin{figure*}
    \centering
    \includegraphics[width=\linewidth]{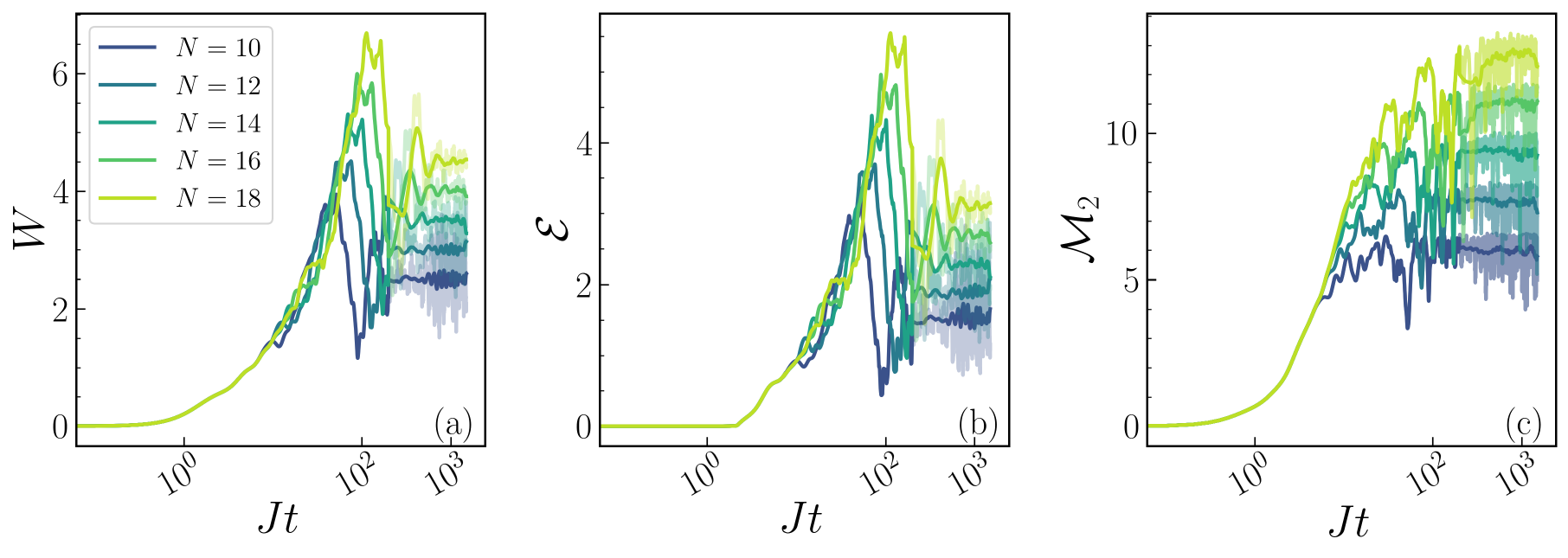}
    \caption{\textbf{Time dynamics in the XXZ model: (a) work (\(W\)) stored in the battery, (b) ergotropy (\(\mathcal{E}\)), and (c) SRE (\(\mathcal{M}_2\)).} The initial growth follows a $t^2$ behavior, as in the two-qubit system. Different shades (lighter to darker) correspond to different system sizes as indicated in the figure. Shaded curves represent the actual values, while solid curves represent time-averaged values. All axes are dimensionless.}
    \label{fig:ergo_work}
\end{figure*}

As mentioned above, the two-qubit system describes the initial dynamics well even for the full XXZ model starting from the initial state, \eqref{inistate}, up to times of the order of $t = 2/J$. Indeed, this is verified in Fig.~\ref{fig:ergo_work}. The time dynamics plotted is obtained using the Chebyshev propagation scheme~\cite{Fehske08}, and a growth of work $W$ quadratic in time, as anticipated from Eq.~\eqref{mwork}, is observed at early times. Similarly, zero ergotropy is observed in this initial window. Physically, the early-time dynamics is dominated by the flip-flop interaction acting on the central bond. This process generates a coherent superposition of the states $\ket{\uparrow\downarrow}$ and $\ket{\downarrow\uparrow}$, as described in Eq.~(\ref{eq:effective_evolution}). A nonzero ergotropy emerges only at higher orders in time, when additional configurations at the center of the chain, such as $\ket{\uparrow\downarrow\downarrow\uparrow}$ and $\ket{\downarrow\uparrow\uparrow\downarrow}$, become populated. These processes occur with a probability of order $\mathcal{O}(J^4 t^4)$, indicating that ergotropy grows more slowly compared to both energy transfer and nonstabilizerness.

\begin{figure}
    \centering
    \includegraphics[width=1.0\linewidth]{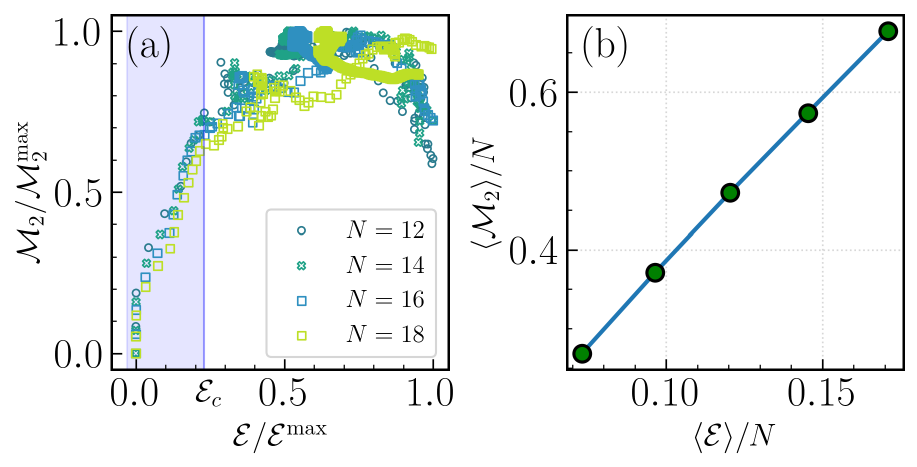}\quad
    \caption{\textbf{SRE against ergotropy.} (a) SRE versus ergotropy (rescaled by their maximal values -- see text) for different system sizes. $\mathcal{E}_c$ indicates the border of the simple, system-size-independent correlation between both observables. (b) Averaged densities of SRE versus ergotropy density, showing that storing a larger average ergotropy requires a larger amount of average SRE.}
    \label{fig:avg_magic_ergotropy}
\end{figure}

At short times, the evolution of the total state as well as the observables is well approximated by the two-qubit system. In particular, a simple Taylor expansion gives for the work $W \approx J^2t^2/4 + \mathcal{O}(t^4)$, while the quantum magic grows as 
\begin{equation}
    \mathcal{M}_2(t)\approx  \frac{J^2t^2}{\ln 2}+ \mathcal{O}(t^4)
\end{equation}
A qualitative relation between the stored energy and nonstabilizerness, given by \(W(t) \sim \mathcal{M}_2(t)\), interestingly extends also to longer times, as shown by the numerical results in Fig.~\ref{fig:ergo_work}. Both observables show significant growth following an initial $t^2$ dependence, then saturate, and at a later stage reveal oscillations around a finite long-time value dependent on the system size, $N$. A similar behavior is shared by the ergotropy for times longer than the initial period when the ergotropy vanishes. 

The behavior observed in Fig.~\ref{fig:ergo_work}, in particular the dependence on the system size $N$, can be comfortably explained by recognizing the various phases of the evolution. First, we have two-qubit dynamics as explained above; then, the domain wall starts to melt and nontrivial many-body dynamics spreads from the center obeying the Lieb--Robinson bound. At a time proportional to the system size, the excitations reach the boundary, and the work, ergotropy, and quantum magic reach their maximal growth, and then stabilize at a system-size-dependent time. Further evolution involves subsequent damped oscillations as the system eventually thermalizes. The time dependence of the observables plotted indicates that beyond the short-time interval where ergotropy vanishes, there is a system-size-dependent time window in which strong correlations between $W$, $\mathcal{E}$, and $\mathcal{M}_2$ exist, and all three quantities grow in a very similar power-law fashion. This is the operational time interval in which the battery is charged. The subsequent evolution at later times reveals substantial fluctuations and eventual thermalization.

Naturally, the system size affects the absolute values of the observables. This effect may be partially eliminated by rescaling the observables (e.g., ergotropy and SRE by their maximal values during the time evolution, $\mathcal{E}^{\rm max}$ and $\mathcal{M}^{\rm max}_2(t)$) and plotting the values parametrically for different times -- compare Fig.~\ref{fig:avg_magic_ergotropy}(a). Specifically, for each time $t$, we numerically evaluate both quantities and construct a parametric plot of scaled SRE versus scaled ergotropy. One observes three regimes: an initial one where SRE grows while ergotropy vanishes; an approximate linear correlation between ergotropy and global nonstabilizerness in the regime of initial growth; and the saturation regime. Clearly, no single functional relation between ergotropy and SRE is valid across the entire time interval. On the other hand, for intermediate times where the initial domain wall spreads across the system leading to the growth of both ergotropy and SRE (as well as the work stored in the battery), such a correlation exists. This region ends when the first maximum is reached, defining the value of scaled $\mathcal{E}_c$ indicated in the figure. 

For longer times, all observables oscillate around their long-time, system-dependent asymptotes. Due to rapid oscillatory behavior, it is more enlightening to examine the time-averaged quantities. To avoid any ambiguity related to the choice of the initial time for the average, we use Eq.~(\ref{eq:avg_figs_of_merit}) and consider the entire time interval (as the initial growth has a negligible effect on the full time average). In Fig.~\ref{fig:avg_magic_ergotropy}(b), we present the average SRE density as a function of the average ergotropy density for different system sizes. Remarkably, we find a simple sublinear dependence between these quantities, indicating a ``no-free-lunch'' principle: on average, storing a larger amount of ergotropy necessarily requires a proportionally larger amount of nonstabilizerness. In other words, average higher ergotropy comes at the cost of increased SRE.

\begin{figure*}
    \centering
    \includegraphics[width=0.9\linewidth]{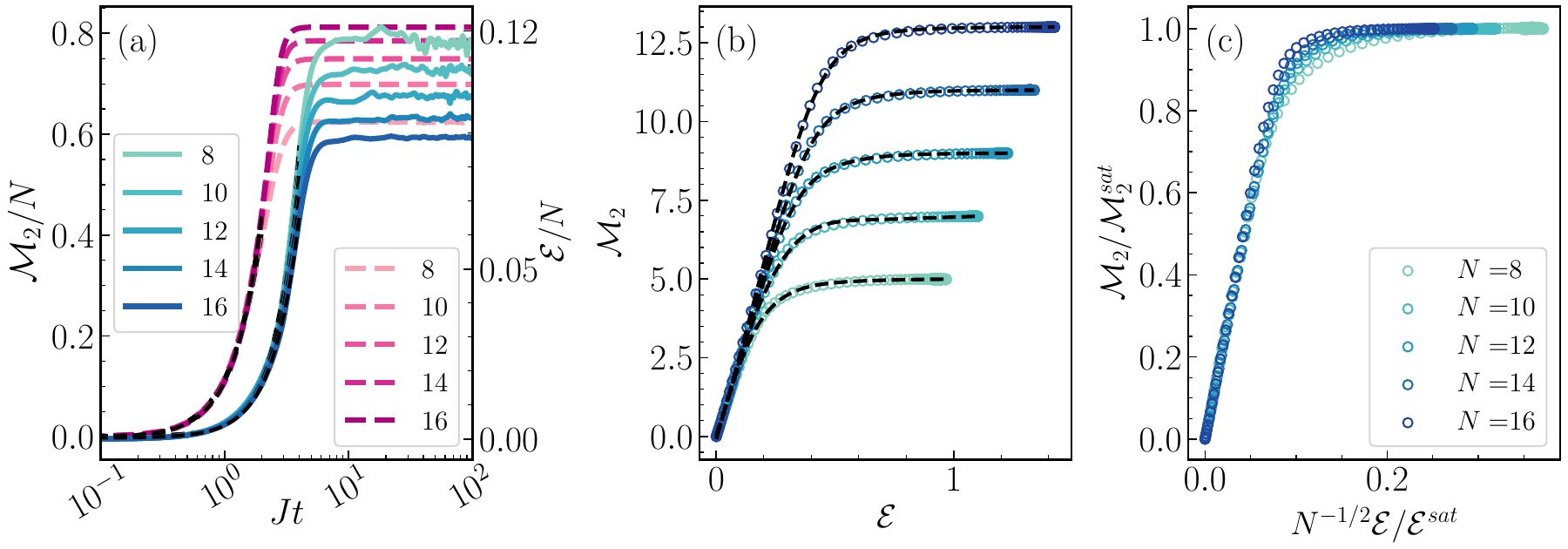}
    \caption{\textbf{Dynamics of charging in the cSYK model.} (a) Time dynamics of SRE (left axis, dashed curves) and ergotropy (right axis, solid curves) for different system sizes; (b) SRE against ergotropy [data from panel (a); dashed lines represent a fit to expression \eqref{tanfit}]; (c) Collapse of data (rescaled by long-time values) across different system sizes. All axes are dimensionless.}
    \label{fig:magic_SYK}
\end{figure*}

\section{Battery charged via complex-SYK interaction}
\label{sec:syk_battery}
In addition to local interactions, we consider an all-to-all interacting Sachdev--Ye--Kitaev (SYK) model~\cite{sachdev_prl_1993}, which is well known for exhibiting fast scrambling dynamics and quantum advantage. The SYK Hamiltonian is given by
\begin{equation}
    H_{BC}^{\mathrm{cSYK}} 
    = \frac{1}{\sqrt{N^3}}\sum_{i<j<k<l}^{N} J_{ijkl}\, \hat{c}_i^\dagger \hat{c}_j^\dagger \hat{c}_k \hat{c}_l,
    \label{eq:hamSYK}
\end{equation}
where $\hat{c}_j^\dagger$ and $\hat{c}_j$ denote fermionic creation and annihilation operators, respectively. This model can be mapped onto a spin-$\tfrac{1}{2}$ system through the Jordan--Wigner transformation,
\begin{equation}
    \hat{c}_j^\dagger 
    = \hat{\sigma}_j^{+} \prod_{m=1}^{j-1} \hat{\sigma}_m^{z}, 
    \qquad
    \hat{\sigma}_j^{\pm} 
    = \frac{1}{2} \left( \hat{\sigma}_j^{x} \pm i \hat{\sigma}_j^{y} \right).
\end{equation}
The coupling coefficients $J_{ijkl}$ are independent Gaussian-distributed complex random variables with mean \(\langle\!\langle J_{ijkl}\rangle\!\rangle=0\) and variance \(\langle\!\langle  \abs{J_{ijkl}}^2\rangle\!\rangle=J^2\), with \(J\in \mathbb{R}\). In addition, the couplings satisfy the symmetry relations \(J_{ijkl}=-J_{jikl}=J_{ijlk}=J_{klij}^*\), which ensure that the Hamiltonian is \(\mathbb{U}(1)\)-symmetric, i.e., \([H_{BC}^{\mathrm{cSYK}}, \sum_{i=1}^N\sigma_z^i]=0\). As a result, the total magnetization, and therefore the total energy of the charger-battery setup, remains conserved. To account for disorder, we consider ensemble-averaged observables. For any observable $O$, the disorder average is defined as
\begin{equation}
    \langle\!\langle O \rangle\!\rangle 
    = \int P(\{J_{ijkl}\})\, O(\{J_{ijkl}\})\, d\{J_{ijkl}\},
\end{equation}
where \(O\) corresponds to the SRE (\(\mathcal{M}_2\)), stored work (\(W\)), or ergotropy (\(\mathcal{E}\)). Such models have recently been explored in the context of SRE spreading~\cite{bera_scipost_2025,russomanno_prb_2025,jasser_prb_2025} and ultra-stable battery construction~\cite{rossini_prl_2020,Rosa2020}, owing to the maximal scrambling nature of the SYK model. Here, our focus is to uncover the relationship between ergotropy and global SRE.

In analogy with previous studies, we find that the SRE exhibits a power-law growth with time, \(\mathcal{M}_2\sim t^\alpha\) with \(\alpha\approx 2\), before saturating to a finite value (see the left axis of Fig.~\ref{fig:magic_SYK}(a))~\cite{bera_scipost_2025}. Importantly, this saturation value differs from the Haar-random limit because the \(\mathbb{U}(1)\)-symmetric \(\mathrm{cSYK}\) dynamics explores only a symmetry-constrained subspace of the total Hilbert space. Similarly, the ergotropy also grows as a power law, \(\sim t^\beta\) with \(\beta\approx 2\), and saturates to a finite steady-state value \(\mathcal{E}^{sat}\) (see the right axis of Fig.~\ref{fig:magic_SYK}(a)). Physically, the \(\mathrm{cSYK}\) model represents a long-range, all-to-all interacting system that simultaneously facilitates energy exchange between the charger and the battery while generating coherence in the battery’s energy eigenbasis. As a consequence, unlike the \(\mathrm{XXZ}\) model, the battery develops a nonzero ergotropy already at short times. This indicates that, under \(\mathrm{cSYK}\) dynamics, both ergotropy and SRE exhibit qualitatively identical growth behavior. Owing to the \(\mathbb{U}(1)\) symmetry, the reduced state of the battery does not evolve toward a maximally mixed state (the so-called Haar state). Instead, its steady-state form can be written as
\begin{equation}
    \rho_B^{sat}\approx\bigoplus_{m}p(m)\frac{\mathbb{I}_{m}}{d_{m}},
    \label{eq:battery_state}
\end{equation}
where \(p_m= \frac{ \binom{n_b}{n_b/2+m} \binom{n_b} {n_b/2-m} }{ \binom{N}{N/2} }\) gives the weight in the magnetization sector \(m\), and \(d_{m}\) denotes the dimension of that sector. The corresponding value of $\mathcal{M}_2$ is available analytically~\cite{iannotti2026}, albeit in a complicated form not reproduced here. It is quite close to the simple approximation
\begin{equation}
   \mathcal{M}_2^{sat}\approx\log_2\binom{N}{N/2}.
   \label{eq: tan}
\end{equation}
A state of the form \eqref{eq:battery_state} is not passive, allowing for a finite steady-state ergotropy. By careful inspection of Fig.~\ref{fig:magic_SYK}(a), one may notice that while the ergotropy density \(\mathcal{E}^{sat}/N\) decreases with increasing system size, the SRE density \(\mathcal{M}_2^{sat}/N\) increases, indicating an inverse trend between the two densities.

To further elucidate the connection between magic growth and ergotropy storage, we plot SRE as a function of ergotropy in Fig.~\ref{fig:magic_SYK}(b). At short times, the SRE grows linearly with ergotropy, independent of system size, demonstrating a one-to-one relation between the two quantities. This behavior clearly reflects a ``no-free-lunch'' principle: storing more ergotropy requires a larger amount of SRE. At later times, however, the SRE gradually saturates while ergotropy continues to increase [compare panel (a)]. In this regime, the two resources behave more independently, and additional ergotropy can be stored without a proportional increase in nonstabilizerness. The growth of nonstabilizerness primarily originates from the generation of coherence in the Pauli basis, whereas the increase of ergotropy is driven by both the transfer of energy from the charger to the battery and the buildup of coherence in the battery’s energy eigenbasis. 

An alternative explanation of the delay between the saturation of magic and ergotropy comes from the fact that the former is entirely determined by the dynamics of the initial state, which lies roughly in the middle of the spectrum of the cSYK Hamiltonian \eqref{eq:hamSYK} and rapidly thermalizes. On the other hand, the passive state formation takes much longer, as that state is necessarily located low in energy, in a region with smaller density of states and slower dynamics.

One is tempted to fit the numerical data to an analytic formula. Such an attempt reveals a hyperbolic scaling relation of the form
\begin{equation}
    \mathcal{M}_2\sim A\tanh(B\mathcal{E})+C\tanh(D\mathcal{E}^2),
    \label{tanfit}
\end{equation}
where \(A\), \(B\), \(C\), and \(D\) are fitting parameters that depend on system size.  Such a fitting function is chosen based on the observation that both the SRE and ergotropy show first a quadratic scaling with time, whereas the saturation time scales differently.  This scaling captures the quantitative relation between SRE and ergotropy in a \(\mathrm{cSYK}\)-charged quantum battery, as shown in Fig.~\ref{fig:magic_SYK}(b) by dashed lines. Rescaling both ergotropy and SRE by their saturation, long-time values yields a single system-size-independent relation, provided the ergotropy is additionally rescaled by a factor of $1/\sqrt{N}$ as seen in Fig.~\ref{fig:magic_SYK}(c). This is related to the general form of the long-time density matrix for $\mathbb{U}(1)$ models undergoing thermalization, as discussed in the next section and argued analytically in Appendix~\ref{app:u1_ergotropy}.

\section{Nonstabilizerness and ergotropy in brick-wall circuits}
\label{sec:magic_brick_wall}
To understand the interplay between stored energy and nonstabilizerness in a generic many-body setting, we consider one of the most minimal yet powerful circuit architectures: the \emph{brick-wall} random unitary circuit (RUC) (see Fig.~\ref{fig:schematics}). The time-evolution operator after circuit depth $t$ (interpreted as discrete time) is given by $U_t = \prod_{l=1}^{t} U^{(l)}$, where $U^{(l)}$ denotes the $l$-th layer of the circuit. The circuit layers alternate between two fixed nearest-neighbor patterns,
\begin{equation}
U^{(2m)} = \prod_{i=1}^{N/2-1} u_{2i,\,2i+1}, 
\qquad
U^{(2m+1)} = \prod_{i=1}^{N/2} u_{2i-1,\,2i},
\label{eq:brickwall}
\end{equation}
with $m \in \mathbb{N}$. Investigating quantum batteries within such circuit models falls naturally into the framework of random quantum batteries~\cite{caravelli_prr_2020,imai_pra_2023,sarkar2025} and sheds light on the interplay between nonstabilizerness and ergodic dynamics~\cite{turkeshi_prb_2025,tirrito_prl_2025,dowling_prl_2025,bejan2025,maity_prr_2026}. Within this framework, our goal is to uncover the relation between the work stored in the battery and the growth of nonstabilizerness. The battery Hamiltonian is defined as in Eq.~(\ref{eq:battery_hamiltonian}). The initial state of the composite charger-battery setup is again chosen as the domain-wall state in Eq.~\eqref{inistate}. To systematically examine how different interaction structures affect the charging dynamics, each two-site gate $u_{i,j}$ acting on neighboring sites $i$ and $j$ is independently drawn from one of several distinct unitary ensembles:
\begin{itemize}
    \item \textbf{Symmetry-constrained evolution:} $u_{i,i+1} \in \mathbb{U}(1)$, corresponding to Haar-random unitaries that preserve total magnetization.
    \item \textbf{Hamiltonian-generated gates:} $u_{i,i+1} = e^{-i H_{ij} \tau}$, where each two-qubit gate is generated by a physically motivated microscopic Hamiltonian.
    \item \textbf{Fully chaotic dynamics:} $u_{i,i+1} \in \mathrm{Haar}$, corresponding to generic Haar-random unitaries acting on the full four-dimensional two-qubit Hilbert space.
    \item \textbf{Clifford circuits:} $u_{i,i+1} \in \mathcal{C}_N$, representing efficiently simulable stabilizer dynamics.
\end{itemize}

\begin{figure}
    \centering
    \includegraphics[width=\linewidth]{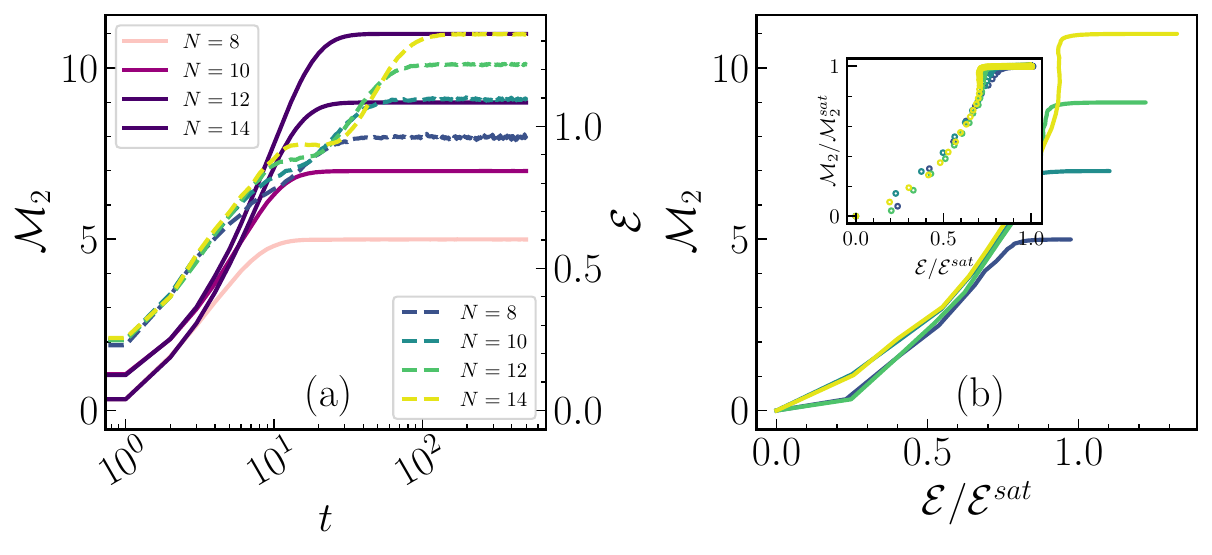}
    \caption{\textbf{SRE and ergotropy for a $\mathbb{U}(1)$-symmetric brick-wall circuit.} (a) SRE (left axis, solid lines) and ergotropy (right axis, dashed lines) as functions of circuit depth $t$. Both quantities saturate to finite steady-state values, matching the qualitative behavior observed in the cSYK model [Fig.~\ref{fig:magic_SYK}(a)]. However, their early-time growth rates differ from those of the cSYK model due to the spatial locality of the circuit. (b) SRE is plotted against normalized ergotropy. A one-to-one correspondence between SRE and ergotropy emerges at early times before $\mathcal{M}_2$ saturates (see inset). All quantities are dimensionless.}
    \label{fig:U1_random}
\end{figure}

The state of the composite system after evolution is given by $\ket{\psi_t} = U_t \ket{\psi(0)}$, and the work stored in the battery can be expressed as
\begin{align}
    W(t) &= \Tr[H_B \ket{\psi_t}\bra{\psi_t}] - \Tr[H_B \ket{\psi_B}\bra{\psi_B}] \nonumber\\
         &= \frac{n_b}{2} + \sum_{i=1}^{n_b} \Tr[\rho^i_B(t) S_z^i] 
          = \frac{1}{2}\left( n_b + \sum_{i=1}^{n_b} m_z^i(t) \right),
\end{align}
where $\rho^i_B(t)$ is the single-site reduced density matrix of the $i$-th battery spin and $m_z^i(t) = \langle \sigma_z^i(t) \rangle$ denotes the corresponding local magnetization. The ergotropy stored in the battery is given by
\begin{equation}
    \mathcal{E}(t) = \Tr[\rho_B(t) H_B] - \min_{U_e} \Tr[U_e \rho_B(t) U_e^\dagger H_B],
\end{equation}
where the minimization is performed over all unitary operators $U_e$ acting on the battery Hilbert space, and $\rho_B(t) = \Tr_C [\ket{\psi_t}\bra{\psi_t}]$ is the reduced state of the battery.

\subsection{$\mathbb{U}(1)$-symmetric unitaries}
The results presented in Secs.~\ref{sec:xxz_battery} and \ref{sec:syk_battery} indicate that a well-defined correspondence between global SRE and stored ergotropy emerges whenever the charging Hamiltonian preserves $\mathbb{U}(1)$ symmetry. Motivated by this observation, we examine RUCs where each two-qubit gate strictly respects $\mathbb{U}(1)$ symmetry~\cite{agarwal_prx_2022,chakraborty_prb_2024,Summer2026,Li2026_b}. In this setting, the total energy of the composite system is conserved, and the battery is charged entirely through energy transferred from the charger.

In Fig.~\ref{fig:U1_random}(a), we plot the evolution of SRE and ergotropy as functions of circuit depth $t$. The SRE grows more slowly than in the $\mathrm{cSYK}$ model. This difference stems directly from the locality of the circuit: the $\mathrm{cSYK}$ Hamiltonian features all-to-all interactions that rapidly scramble nonstabilizerness across the entire system, whereas the brick-wall architecture is constrained by local light-cone spreading. Nevertheless, the SRE saturates to a value consistent with the analytical estimate in Eq.~\eqref{eq: tan}, with the agreement improving systematically with increasing system size. In parallel, the ergotropy initially grows as $\mathcal{E}(t) \propto t^\beta$ with $\beta \approx 2$ before saturating at $\mathcal{E}^{\mathrm{sat}}$, matching the values found in the $\mathrm{cSYK}$ model. The steady-state ergotropy scales as $\mathcal{E}^{\mathrm{sat}} \propto \sqrt{N}$, which can be deduced from the block-diagonal structure of the thermalized battery state in Eq.~(\ref{eq:battery_state}). Because the associated passive state concentrates its largest weights in the vicinity of the $m=0$ magnetization sector, the passive-state energy is determined by fluctuations around this sector (see Appendix~\ref{app:u1_ergotropy} for details).

The functional dependence of SRE on ergotropy differs from that of the $\mathrm{cSYK}$ model. As shown in the inset of Fig.~\ref{fig:U1_random}(b), the SRE initially grows slowly with ergotropy, indicating that storing small amounts of extractable work requires only a minor nonstabilizerness cost. At intermediate ergotropy values, $\mathcal{M}_2$ increases sharply before reaching saturation. Consequently, once this saturation regime is reached, the battery can continue to accumulate ergotropy without requiring further nonstabilizerness generation.

\begin{figure*}[t]
    \centering
    \includegraphics[width=\linewidth]{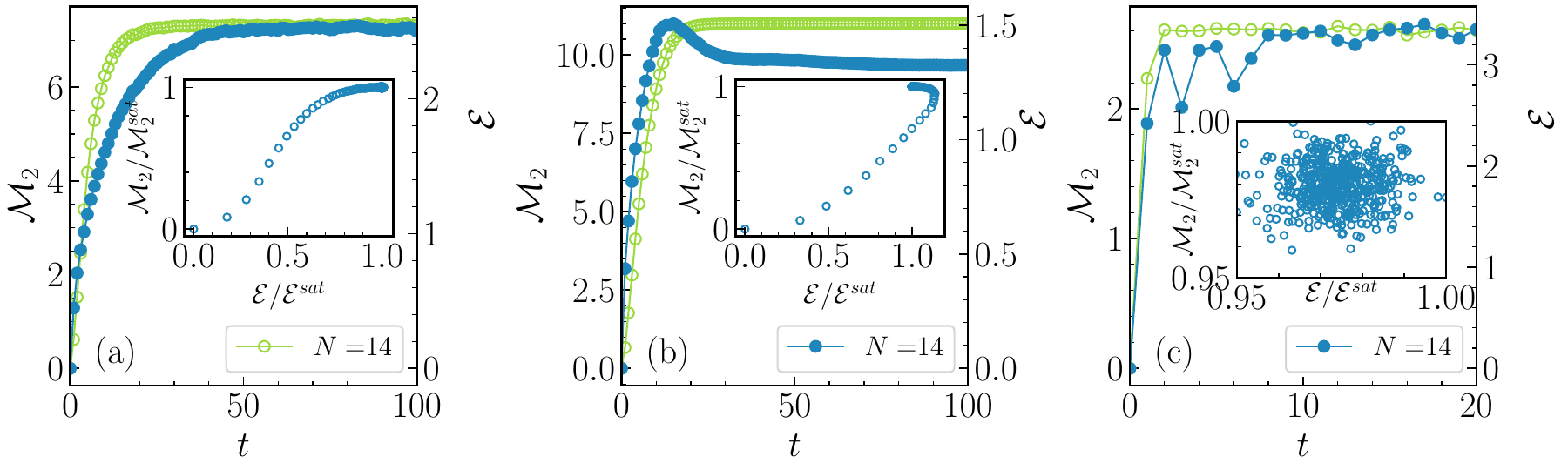}
    \caption{\textbf{SRE and ergotropy for brick-wall circuits generated by specific physical Hamiltonians.} Time evolution of SRE (left axes) and ergotropy (right axes) for (a) $H_{BC}^{XX}$, (b) $H_{BC}^{XXZ}$, and (c) \(H_{BC}^{Is}\).  The insets illustrate that a  functional correspondence between nonstabilizerness and ergotropy is observed for both the $H_{BC}^{XX}$ and $H_{BC}^{XXZ}$ models, whereas it is completely absent for Ising interactions. All quantities are dimensionless.}
    \label{fig:random_hamiltonian}
\end{figure*}

\subsection{Hamiltonian-generated gates}
To bridge the gap between abstract circuit ensembles and experimental implementations, we investigate two-qubit unitaries generated by microscopic physical Hamiltonians. Specifically, we study three classes of nearest-neighbor couplings:
\begin{align}
    H_{BC}^{\mathrm{Is}} &= J_{ij} \sigma_x^i \sigma_x^j, \nonumber\\
    H_{BC}^{XX} &= J_{ij} (\sigma_x^i \sigma_x^j + \sigma_y^i \sigma_y^j), \nonumber\\
    H_{BC}^{XXZ} &= J_{ij} (\sigma_x^i \sigma_x^j + \sigma_y^i \sigma_y^j + \Delta \sigma_z^i \sigma_z^j),
\end{align}
where the coupling strengths $J_{ij}$ are drawn independently from a uniform distribution $J_{ij} \in [0, J]$. These models possess distinct symmetry structures: $H_{BC}^{\mathrm{Is}}$ preserves a $\mathbb{Z}_2$ parity symmetry, whereas both $H_{BC}^{XX}$ and $H_{BC}^{XXZ}$ preserve $\mathbb{U}(1)$ magnetization symmetry. The additional $ZZ$ interaction in $H_{BC}^{XXZ}$ further introduces many-body scrambling. For all of our calculation we put \(J=1\).

The early-time growth of both SRE and ergotropy is qualitatively similar for the $H_{BC}^{XX}$ and $H_{BC}^{XXZ}$ interactions, following the characteristic behavior of generic $\mathbb{U}(1)$-symmetric circuits. However, their intermediate-time dynamics diverge markedly. For the $H_{BC}^{XX}$ interaction, both SRE and ergotropy grow monotonically toward saturation [Fig.~\ref{fig:random_hamiltonian}(a)]. In contrast, for the $H_{BC}^{XXZ}$ interaction, the ergotropy reaches a transient maximum and subsequently decreases before stabilizing, while the SRE grows monotonically throughout the entire evolution [Fig.~\ref{fig:random_hamiltonian}(b)]. This demonstrates that even under $\mathbb{U}(1)$-symmetric dynamics, ergotropy can degrade while nonstabilizerness remains constant or increases.

An entirely different picture emerges for the Ising interaction $H_c^{\mathrm{Is}}$, shown in Fig.~\ref{fig:random_hamiltonian}(c). Here, no deterministic functional relationship between SRE and ergotropy is preserved; instead, SRE values scatter widely for any given value of ergotropy. This decoupling confirms that nonstabilizerness and ergotropy evolve independently in the absence of continuous symmetries. From the viewpoint of battery performance, however, the Ising interaction offers a distinct advantage: it yields a significantly higher stored ergotropy while generating the lowest SRE among the three models considered. Even though both $H_c^{XX}$ and $H_c^{\mathrm{Is}}$ map to noninteracting fermions, the Ising interaction provides superior charging performance at a substantially reduced nonstabilizerness cost.

\subsection{Clifford unitaries}
We now turn to circuits composed exclusively of Clifford gates. As formulated in Eqs.~\eqref{eq:battery_hamiltonian} and \eqref{eq:initial_state}, both the battery Hamiltonian and the initial domain-wall state belong to the stabilizer formalism. The state evolves as $\ket{\psi_t} = \mathcal{U}_c \ket{\psi(0)}$, where $\mathcal{U}_c$ is drawn from a brick-wall Clifford circuit~\cite{bera_pra_2020}. Because Clifford unitaries map stabilizer states to stabilizer states, the composite state remains strictly inside the stabilizer manifold at all times, ensuring $\mathcal{M}_2(t) = 0$ identically. 

\begin{figure}[t]
    \centering
    \includegraphics[width=\linewidth]{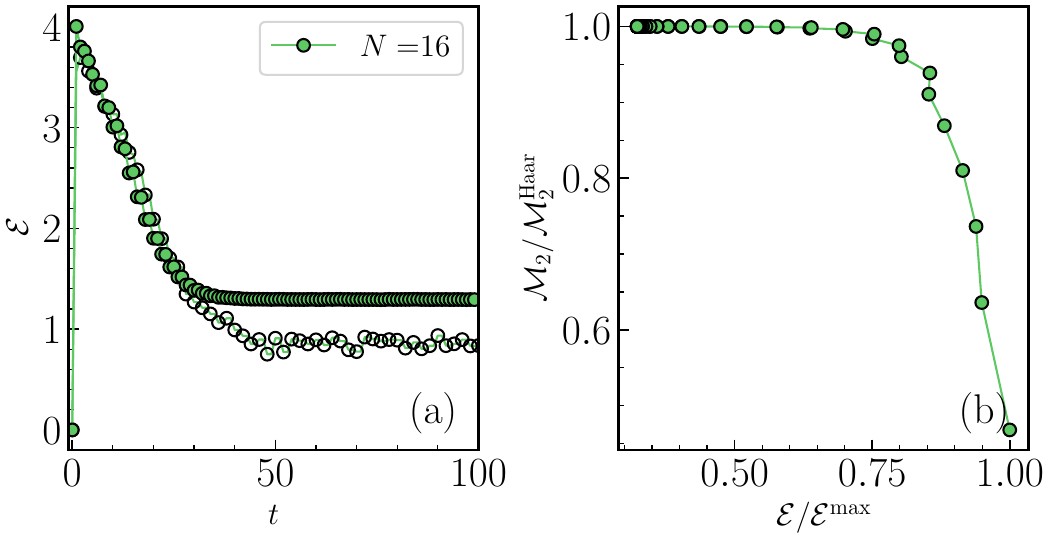}
    \caption{\textbf{Ergotropy and SRE for Haar-random and Clifford brick-wall circuits.} (a) Ergotropy of the battery for Clifford (open circles) and Haar-random (solid circles) circuits, showing identical qualitative behavior. (b) SRE as a function of ergotropy for Haar-random circuits, illustrating that SRE grows while ergotropy decreases. All quantities are dimensionless.}
    \label{fig:haar_random_erg}
\end{figure}

Nonetheless, the evolution of ergotropy remains highly nontrivial [Fig.~\ref{fig:haar_random_erg}(a)]. This demonstrates that finite ergotropy can be stored without generating nonstabilizerness. Immediately after the first layer of gates, the ergotropy undergoes an abrupt jump, followed by a monotonic decrease toward a finite plateau. Concurrently, the stored work saturates to $W = n_b/2 = N/4$, reflecting the fact that the Clifford ensemble forms a unitary $2$-design that efficiently depolarizes local observables toward $\rho_i \approx \mathbb{I}/2$.

To understand the sharp initial surge in ergotropy, we examine the spectral properties of the reduced state under stabilizer dynamics. Tracing out the charger from a global stabilizer state yields a uniform statistical mixture over an affine subspace of the Pauli group~\cite{nielsen2010}. The reduced density matrix of the battery is therefore maximally mixed over its support subspace:
\begin{equation}
    \rho_{n_b}(t) = \frac{1}{2^{n_b - r(t)}} \Pi_B(t),
    \label{eq:reduced_state}
\end{equation}
where $\Pi_B(t)$ is an orthogonal projector whose operator rank (the support dimension of $\rho_{n_b}$) is given by $\mathrm{rank}( \Pi_B(t))=\Tr[\Pi_B(t)] = 2^{n_b - r(t)}$. Here, $r(t) \in \{0, \dots, n_b\}$ is the algebraic rank of the local stabilizer group $\mathcal{S}_B(t) = \mathcal{S}_{\text{total}}(t) \cap \mathcal{P}_{n_b}$ (i.e. the number of independent, commuting Pauli generators supported entirely on the battery subsystem)~\cite{gottesman1997}.  Consequently, the entanglement entropy of the battery is simply $S=n_b - r(t)$ and the spectrum of $\rho_{n_b}(t)$ is flat:
\begin{equation}
    \lambda = 
    \begin{cases} 
        2^{-(n_b - r(t))}, & \text{with multiplicity } 2^{n_b - r(t)}, \\
        0, & \text{with multiplicity } 2^{n_b} - 2^{n_b - r(t)}.
    \end{cases}
    \label{eq:eigenvalue_state}
\end{equation}

For the battery Hamiltonian $H_B = \sum_{i=1}^{n_b} S_z^i$, the distinct many-body energy eigenvalues arranged in ascending order are
\begin{equation}
    E_k = -\frac{n_b}{2} + k, \quad k \in \{0, 1, \dots, n_b\},
\end{equation}
with degeneracies $g_k = \binom{n_b}{k}$, where $k$ counts the number of spin-up excitations.

The passive state $\sigma_{\rho}$ associated with $\rho_{n_b}(t)$ is obtained by pairing the eigenvalues of $\rho_{n_b}(t)$ in descending order with the ascending energy spectrum of $H_B$~\cite{Pusz78}. Because the nonzero spectrum of $\rho_{n_b}(t)$ is completely flat, constructing the passive state corresponds to populating the lowest energy manifolds up to a cutoff level $k^*$, defined by
\begin{equation}
    \sum_{k=0}^{k^*-1} g_k < 2^{n_b - r(t)} \le \sum_{k=0}^{k^*} g_k.
    \label{eq:cut_off}
\end{equation}
Evaluating the passive-state energy $E_{\mathrm{p}}(t) = \Tr[\sigma_\rho H_B]$ yields
\begin{equation}
    E_{\mathrm{p}}(t) = \frac{1}{2^{n_b - r(t)}} \left[ \sum_{k=0}^{k^*-1} g_k E_k + \left( 2^{n_b - r(t)} - \sum_{k=0}^{k^*-1} g_k \right) E_{k^*} \right].
    \label{eq:ergo_stabilizer}
\end{equation}
The ergotropy is then given by
\begin{equation}
    \mathcal{E}(\rho_{n_b}(t)) = \langle H_B(t) \rangle - E_{\mathrm{p}}(t) = \frac{1}{2}\sum_{i=1}^{n_b} m_z^i(t) - E_{\mathrm{p}}(t).
    \label{eq:ergotropy}
\end{equation}

At $t = 0$, the battery is fully polarized ($r(0) = n_b$, $m_z^i = -1$), so $\langle H_B(0) \rangle = E_{\mathrm{p}}(0) = -n_b/2$, giving $\mathcal{E} = 0$. Under the first entangling Clifford layer, local stabilizer operators are rapidly scrambled across the interface, causing $r(t)$ to drop. However, the support dimension $2^{n_b - r(t)}$ initially remains small enough that $k^* \ll n_b$, keeping the passive-state energy pinned near the ground-state floor: $E_{\mathrm{p}}(t) \approx -n_b/2 + \mathcal{O}(1)$. Conversely, Clifford scrambling depolarizes the local magnetization rapidly ($\langle H_B \rangle \to 0$). The disparity between the fast decay of local magnetization and the delayed rise of the passive-state energy generates an extensive jump in extractable ergotropy after a single layer.

At late times, the ergotropy of the battery is fully determined by the asymptotic number of preserved stabilizer generators, as formalized below:

\textcolor{black}{\begin{proposition}
In the thermodynamic limit ($n_b \gg 1$) and at long times, the steady-state ergotropy $\mathcal{E}_\infty$ of a reduced stabilizer state $\rho_B$ is bounded by its asymptotic stabilizer group rank $r_\infty$:
\begin{equation}
    \mathcal{E}_{\infty} \le \frac{n_b}{2}\left[1 - 2 \Gamma^{-1}\left(1 - \frac{r_{\infty}}{n_b}\right)\right],
\end{equation}
where $\Gamma(x) = -x\log_2 x - (1-x)\log_2(1-x)$ is the binary entropy function.
\end{proposition}}

\textcolor{black}{\begin{proof}
Assume that $r_\infty/n_b \to \alpha \in (0,1)$ as
$n_b \to \infty$, and define
\begin{equation*}
    x_0 = \Gamma^{-1}\left(1-\frac{r_\infty}{n_b}\right);
    \quad x^* = \frac{k^*}{n_b}.
\end{equation*}
Now, in the large $n_b$ limit, using Stirling's formula we obtain
\begin{equation*}
g_k = \binom{n_b}{k} = \frac{2^{n_b\Gamma(x)}}{\sqrt{2\pi n_b x(1-x)}} \left[1+\mathcal{O}(n_b^{-1})\right],
\end{equation*}
where \(x={k}/{n_b}\). Now, the degeneracies increase geometrically near the cutoff, with \({g_{k^*-1}}/{g_{k^*}}<1\), i.e., the cumulative degeneracy is of the same order as its boundary term. Using the cutoff condition in Eq.~(\ref{eq:cut_off}), we therefore obtain
\begin{equation}
n_b-r_\infty = n_b\Gamma(x^*)-\frac{1}{2}\log_2 n_b +\mathcal{O}(1).
\end{equation}
Since $\Gamma(x_0)=1-r_\infty/n_b$, expanding around $x_0$ yields
\begin{equation}
k^* = n_b x_0+ \frac{\log_2 n_b}{2\Gamma'(x_0)}
    +\mathcal{O}(1),
\end{equation}
with \(\Gamma'(x_0)>0\). Now, using the relation $E_k=E_{k^*}-(k^*-k)$, Eq.~\ref{eq:ergo_stabilizer} can be rewritten as
\begin{equation}
E_{\mathrm p}  = E_{k^*} - \frac{1}{2^{n_b-r_\infty}} \sum_{k<k^*}g_k(k^*-k).
\label{eq:reexpress_ergo}
\end{equation}
The geometric decay of $g_{k^*-\ell}$ with $\ell$, together with the cutoff condition, ensures that the second term is
$\mathcal{O}(1)$. Consequently,
\begin{equation}
E_{\mathrm p} = -\frac{n_b}{2}+n_b x_0 +\frac{\log_2 n_b}{2\Gamma'(x_0)} +\mathcal{O}(1).
\end{equation}
Now the battery energy vanishes at long times, $\langle H_B\rangle=0$, the asymptotic ergotropy is
\begin{equation*}
\mathcal E_\infty =-E_{\mathrm p} = \frac{n_b}{2}(1-2x_0) -\frac{\log_2 n_b}{2\Gamma'(x_0)} +\mathcal{O}(1).
\end{equation*}
Since $\Gamma'(x_0)$ approaches a finite positive constant, the negative logarithmic correction dominates the bounded remainder for sufficiently large $n_b$. Hence,
\begin{equation}
\mathcal E_\infty \leq \frac{n_b}{2} \left[ 1-2\Gamma^{-1} \left(1-\frac{r_\infty}{n_b}\right)\right],
\end{equation}
which concludes the proof.
\end{proof}}


\subsection{Haar-random unitaries}
When each two-qubit gate is independently sampled from the Haar measure on $\mathbb{U}(4)$, the ergotropy dynamics closely tracks that of the Clifford circuit [Fig.~\ref{fig:haar_random_erg}(a)]. Ergotropy exhibits a rapid initial rise, followed by an intermediate decay toward a steady-state value that is slightly higher than in the Clifford case. This matches the intuition developed for Clifford circuits: fast operator scrambling rapidly pushes the system away from passivity before the passive energy floor can catch up. Unlike Clifford dynamics, however, Haar-random gates continuously generate nonstabilizerness, creating a finite SRE cost throughout the charging process.

Figure~\ref{fig:haar_random_erg}(b) reveals that the SRE increases monotonically as ergotropy decays. Consequently, states possessing higher nonstabilizerness can correspond to lower extractable work. This anticorrelation demonstrates that maximizing magic does not automatically enhance battery performance and can even coincide with a degradation of extractable work.

\section{Role of nonstabilizerness in the initial state}
\label{sec:initial_magic}
We now examine the complementary scenario in which the battery is initialized in a state that inherently contains nonstabilizerness, while the charging protocol is implemented via stroboscopic stabilizer unitaries. Our objective is to determine how the initial SRE influences the maximum average charging power.

We consider a quantum battery described by the one-dimensional anisotropic $XYZ$ model in a magnetic field:
\begin{equation}
    H_B = \sum_{i=1}^{N-1} \left[ J_x'S_x^i S_x^{i+1} + J_y'S_y^i S_y^{i+1} + \Delta' S_z^i S_z^{i+1} \right] 
    + {h'} \sum_{i=1}^{N} S_z^i,
\end{equation}
where $h'$ denotes the strength of the external magnetic field, while \(J_x'\), $J_y'$, and \(\Delta'\) characterize the strength of nearest-neighbor interactions. 
For convenience, we introduce the dimensionless parameters $J_y = J_y'/J_x'$, $\Delta = \Delta'/J_x'$, characterizing nearest-neighbor interactions, and $h = h'/J_x'$ as the strength of the external magnetic field.

The battery is initialized in the ground state of $H_B$, which generally carries a finite nonstabilizerness. The charging is driven by a local transverse field:
\begin{equation}
    H_C = 2J_x'\sum_{i=1}^N S_x^i,
\end{equation}
setting $J_x'=1$ below.
While continuous evolution under $H_C$ is non-Clifford, the operator $U(t) = \exp(-i H_C t)$ reduces to an exact Clifford gate at stroboscopic times $t_k = k\pi/4$ ($k \in \mathbb{N}$). We therefore adopt a pulsed charging protocol evaluated at these discrete intervals~\cite{Ghosh2020}. The maximum average charging power is defined as
\begin{equation}
    P_{\max} = \max_{k} \frac{W(k)}{t_k},
\end{equation}
where $W(k)$ is the work stored after the $k$-th pulse.

\begin{figure}[t]
    \centering
    \includegraphics[width=\linewidth]{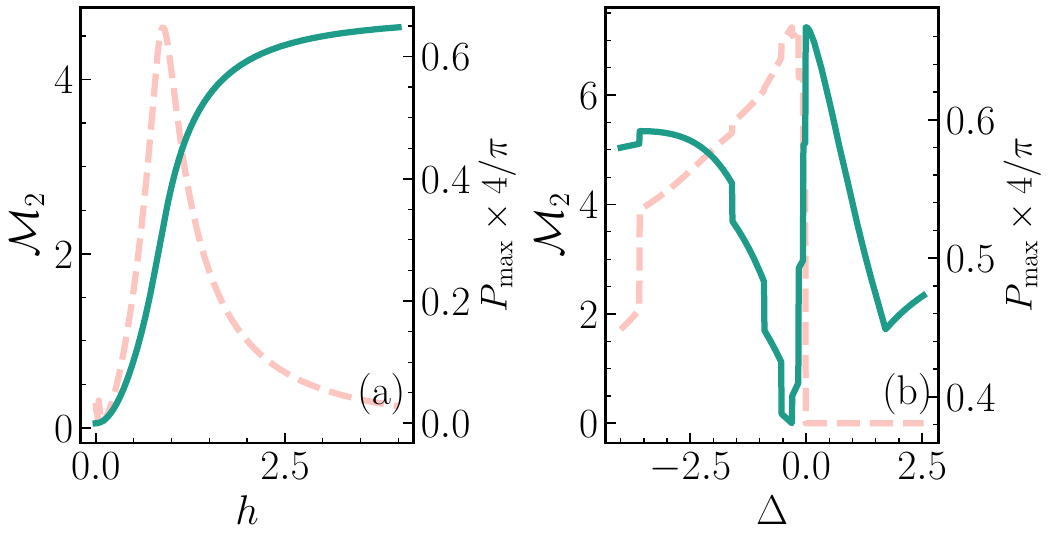}
    \caption{\textbf{SRE of the initial state and the maximum average charging power.} Panel (a): Ising case with \(J_y=\Delta=0\) and \(J_x'=1\). SRE (dashed pink curve) and $P_{\mathrm{max}}$ (solid green curve) are plotted as functions of the magnetic field. Panel (b): XXZ Hamiltonian case with \(J_x'=J_y=1\) and \(h=0.5\) as a function of $\Delta$ with the same colors for SRE and  $P_{\mathrm{max}}$ as in (a). Note that in both cases, higher initial magic does not correlate with enhanced charging power. All quantities are dimensionless.}
    \label{fig:xy_battery_stabilizer}
\end{figure}

We consider two cases. The Ising Hamiltonian is realized for $J_y=\Delta=0$.
Figure~\ref{fig:xy_battery_stabilizer}(a) presents the SRE of the initial state and $P_{\max}$ as functions of the transverse field $h$. The ground-state SRE peaks near the critical point $h_c = 1$ and falls off in the asymptotic phases. In contrast, $P_{\max}$ increases monotonically with $h$ before saturating for large field values. Figure~\ref{fig:xy_battery_stabilizer}(b) corresponds to the $XXZ$ model with $J_x'=J_y=1$. Here, we fix the magnetic field at $h=0.5$ and vary the interaction strength $\Delta$. No correlation between the SRE and the charging power is observed. Both change very rapidly at $\Delta=0$ (which corresponds to a noninteracting limit in the fermionic language after a Jordan--Wigner transformation). The ground state is a stabilizer state for $\Delta>0$ (see also Ref.~\cite{Viscardi26}). Both models therefore provide compelling evidence that, in general, there is no obvious relation between the SRE of the initial state and the maximal charging power. For the Ising model, the highest average charging power is attained in the large-$h$ limit where the initial state has negligible magic. This shows that initial nonstabilizerness does not constitute a generic resource for enhancing charging power and can even act as a bottleneck.

\section{Conclusions}
\label{sec:conclusion}
In this work, we have investigated the role of nonstabilizerness (quantum magic) in the operation of quantum batteries. By modeling the composite charger-battery system as an interacting spin chain, we quantified the nonstabilizerness generated during the charging process using the second-order stabilizer R\'enyi entropy ($\mathcal{M}_2$).

For both nearest-neighbor $\mathrm{XXZ}$ chains and all-to-all complex Sachdev--Ye--Kitaev ($\mathrm{cSYK}$) interactions, the conservation of total magnetization ($\mathbb{U}(1)$ symmetry) establishes a direct correspondence between SRE and the stored ergotropy. In the $\mathrm{cSYK}$ model, this manifests as a robust functional relation between disorder-averaged quantities (SRE and ergotropy) across different system sizes at all times. For other $\mathbb{U}(1)$-symmetric models, such a functional relation is realized for limited times only, typically when both SRE and ergotropy increase at early times. Later, when both quantities attempt to reach their asymptotic long-time values, it appears that ergotropy saturates much more slowly, which may be linked to the slow thermalization of the passive state. In contrast, for non-symmetric models, such as transverse-field Ising circuits or fully generic Haar-random circuits, this correspondence breaks down entirely, with SRE and ergotropy evolving independently.

Crucially, our findings demonstrate that nonstabilizerness is neither necessary nor universally beneficial for quantum battery operation. First, pure Clifford circuits can generate substantial ergotropy without producing any nonstabilizerness ($\mathcal{M}_2 = 0$). Second, in Haar-random dynamics, SRE continues to increase while ergotropy degrades at intermediate times. Finally, when charging nonstabilizer ground states via stroboscopic Clifford pulses, we found that initial magic is nonmonotonically related to charging power, with optimal power achieved in regimes of low initial nonstabilizerness. Together, these results clarify the operational boundary between classical simulability and thermodynamic performance in quantum energy storage devices.

\begin{acknowledgments}
We thank Alessandro d'Oliveira Junior for discussions. This research was carried out and financed within the framework of the second Swiss Contribution MAPS (Grant No.~230870). Partial support by the National Science Centre (Poland) under project 2021/43/I/ST3/01142 (OPUS call within the WEAVE programme) is gratefully acknowledged (J.Z.). We also acknowledge the Polish high-performance computing infrastructure PLGrid (HPC Center: ACK Cyfronet AGH) for computational facilities and support under grant no.~PLG/2025/018400.
\end{acknowledgments}

\appendix

\section{Asymptotic ergotropy scaling under $\mathbb{U}(1)$ conservation}
\label{app:u1_ergotropy}
At late times ($t \gg N^2$), the global state thermalizes toward an effective Haar-random state within the conserved zero-magnetization sector $S^z_{\mathrm{tot}} = 0$. For a system of size $N = 2n_b$, the reduced density matrix of the battery ($n_b$ spins) is block-diagonal with respect to the subsystem magnetization sectors $m \in [-n_b/2, \, n_b/2]$:
\begin{equation}
\rho_B = \frac{1}{\binom{N}{N/2}} \bigoplus_{m=-n_b/2}^{n_b/2} \binom{n_b}{n_b/2-m} \mathbb{I}_{d_m},
\end{equation}
where $d_m = \binom{n_b}{n_b/2+m}$ is the sector dimension, and the total probability weight of block $m$ is given by
\begin{equation}
p_m = \frac{\binom{n_b}{n_b/2+m}\binom{n_b}{n_b/2-m}}{\binom{N}{N/2}}.
\end{equation}
The energy of each sector is $E_m = m$. Because $p_m = p_{-m}$, the mean energy vanishes identically: $\Tr[\rho_B H_B] = \sum_m m \, p_m = 0$.
\begin{figure}[t]
    \centering
    \includegraphics[width=0.7\linewidth]{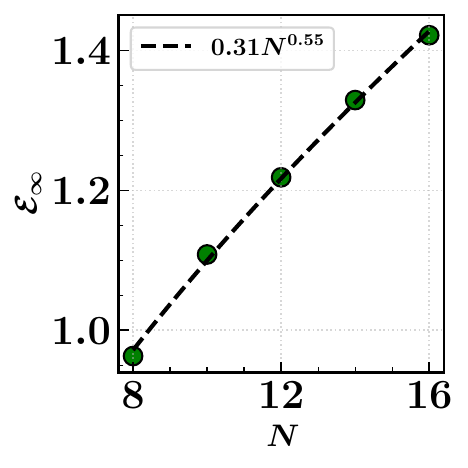}
    \caption{\textbf{Asymptotic ergotropy scaling with system size.} Late-time ergotropy $\mathcal{E}_\infty$ for the $\mathbb{U}(1)$-symmetric brick-wall circuit as a function of system size $N$. A two-parameter fit $a_1 N^{a_2}$ yields $a_2 \approx 0.55$, consistent with the analytical $\sqrt{N}$ scaling. All quantities are dimensionless.}
    \label{fig:u1_ergotropy}
\end{figure}

In the passive state, probabilities are monotonically rearranged against energy. Because the largest weights $p_m$ occur near $m = 0$, the passive configuration effectively maps each sector energy $m \to -\vert{}m\vert{}$ to leading order, giving
\begin{equation}
E_{\mathrm{p}} \approx -\sum_{m} |m| \, p_m = -\mathbb{E}[|m|].
\label{eq:abr_u1_ergo}
\end{equation}
Applying Stirling's approximation for large $N$, the discrete distribution $p_m$ converges to a Gaussian distribution:
\begin{equation}
p_m \approx \frac{1}{\sqrt{2\pi\sigma^2}} \exp\left(-\frac{m^2}{2\sigma^2}\right), \qquad \sigma^2 = \frac{n_b}{4} = \frac{N}{8}.
\end{equation}
For a centered Gaussian variable, the first absolute moment evaluates to $\mathbb{E}[\vert{}m\vert{}] = \sigma \sqrt{2/\pi}$. Substituting $\sigma = \sqrt{N/8}$ yields
\begin{equation}
E_{\mathrm{p}} \approx -\sqrt{\frac{N}{4\pi}}.
\end{equation}
Because the mean energy is zero, the asymptotic ergotropy scales subextensively as
\begin{equation}
\mathcal{E}_{\infty} = \langle H_B \rangle - E_{\mathrm{p}} \approx \sqrt{\frac{N}{4\pi}} \sim \sqrt{N}.
\end{equation}

To confirm this analytical prediction, we plot $\mathcal{E}_\infty$ as a function of $N$ in Fig.~\ref{fig:u1_ergotropy}. A power-law fit $\mathcal{E}_\infty = a_1 N^{a_2}$ confirms the subextensive scaling exponent $a_2 \approx 0.5$.

\bibliography{ref}

\begin{thebibliography}{99}%
\makeatletter
\providecommand \@ifxundefined [1]{%
 \@ifx{#1\undefined}
}%
\providecommand \@ifnum [1]{%
 \ifnum #1\expandafter \@firstoftwo
 \else \expandafter \@secondoftwo
 \fi
}%
\providecommand \@ifx [1]{%
 \ifx #1\expandafter \@firstoftwo
 \else \expandafter \@secondoftwo
 \fi
}%
\providecommand \natexlab [1]{#1}%
\providecommand \enquote  [1]{``#1''}%
\providecommand \bibnamefont  [1]{#1}%
\providecommand \bibfnamefont [1]{#1}%
\providecommand \citenamefont [1]{#1}%
\providecommand \href@noop [0]{\@secondoftwo}%
\providecommand \href [0]{\begingroup \@sanitize@url \@href}%
\providecommand \@href[1]{\@@startlink{#1}\@@href}%
\providecommand \@@href[1]{\endgroup#1\@@endlink}%
\providecommand \@sanitize@url [0]{\catcode `\\12\catcode `\$12\catcode
  `\&12\catcode `\#12\catcode `\^12\catcode `\_12\catcode `\%12\relax}%
\providecommand \@@startlink[1]{}%
\providecommand \@@endlink[0]{}%
\providecommand \url  [0]{\begingroup\@sanitize@url \@url }%
\providecommand \@url [1]{\endgroup\@href {#1}{\urlprefix }}%
\providecommand \urlprefix  [0]{URL }%
\providecommand \Eprint [0]{\href }%
\providecommand \doibase [0]{https://doi.org/}%
\providecommand \selectlanguage [0]{\@gobble}%
\providecommand \bibinfo  [0]{\@secondoftwo}%
\providecommand \bibfield  [0]{\@secondoftwo}%
\providecommand \translation [1]{[#1]}%
\providecommand \BibitemOpen [0]{}%
\providecommand \bibitemStop [0]{}%
\providecommand \bibitemNoStop [0]{.\EOS\space}%
\providecommand \EOS [0]{\spacefactor3000\relax}%
\providecommand \BibitemShut  [1]{\csname bibitem#1\endcsname}%
\let\auto@bib@innerbib\@empty
\bibitem [{\citenamefont {Alicki}\ and\ \citenamefont
  {Fannes}(2013)}]{Alicki2013}%
  \BibitemOpen
  \bibfield  {author} {\bibinfo {author} {\bibfnamefont {R.}~\bibnamefont
  {Alicki}}\ and\ \bibinfo {author} {\bibfnamefont {M.}~\bibnamefont
  {Fannes}},\ }\bibfield  {title} {\bibinfo {title} {Entanglement boost for
  extractable work from ensembles of quantum batteries},\ }\href
  {https://doi.org/10.1103/PhysRevE.87.042123} {\bibfield  {journal} {\bibinfo
  {journal} {Phys. Rev. E}\ }\textbf {\bibinfo {volume} {87}},\ \bibinfo
  {pages} {042123} (\bibinfo {year} {2013})}\BibitemShut {NoStop}%
\bibitem [{\citenamefont {Campaioli}\ \emph {et~al.}(2024)\citenamefont
  {Campaioli}, \citenamefont {Gherardini}, \citenamefont {Quach}, \citenamefont
  {Polini},\ and\ \citenamefont {Andolina}}]{quantum_battery_review}%
  \BibitemOpen
  \bibfield  {author} {\bibinfo {author} {\bibfnamefont {F.}~\bibnamefont
  {Campaioli}}, \bibinfo {author} {\bibfnamefont {S.}~\bibnamefont
  {Gherardini}}, \bibinfo {author} {\bibfnamefont {J.~Q.}\ \bibnamefont
  {Quach}}, \bibinfo {author} {\bibfnamefont {M.}~\bibnamefont {Polini}},\ and\
  \bibinfo {author} {\bibfnamefont {G.~M.}\ \bibnamefont {Andolina}},\
  }\bibfield  {title} {\bibinfo {title} {Colloquium: Quantum batteries},\
  }\href {https://doi.org/10.1103/RevModPhys.96.031001} {\bibfield  {journal}
  {\bibinfo  {journal} {Rev. Mod. Phys.}\ }\textbf {\bibinfo {volume} {96}},\
  \bibinfo {pages} {031001} (\bibinfo {year} {2024})}\BibitemShut {NoStop}%
\bibitem [{\citenamefont {Binder}\ \emph {et~al.}(2015)\citenamefont {Binder},
  \citenamefont {Vinjanampathy}, \citenamefont {Modi},\ and\ \citenamefont
  {Goold}}]{Binder2015}%
  \BibitemOpen
  \bibfield  {author} {\bibinfo {author} {\bibfnamefont {F.~C.}\ \bibnamefont
  {Binder}}, \bibinfo {author} {\bibfnamefont {S.}~\bibnamefont
  {Vinjanampathy}}, \bibinfo {author} {\bibfnamefont {K.}~\bibnamefont
  {Modi}},\ and\ \bibinfo {author} {\bibfnamefont {J.}~\bibnamefont {Goold}},\
  }\bibfield  {title} {\bibinfo {title} {Quantacell: powerful charging of
  quantum batteries},\ }\href {https://doi.org/10.1088/1367-2630/17/7/075015}
  {\bibfield  {journal} {\bibinfo  {journal} {New Journal of Physics}\ }\textbf
  {\bibinfo {volume} {17}},\ \bibinfo {pages} {075015} (\bibinfo {year}
  {2015})}\BibitemShut {NoStop}%
\bibitem [{\citenamefont {Campaioli}\ \emph {et~al.}(2017)\citenamefont
  {Campaioli}, \citenamefont {Pollock}, \citenamefont {Binder}, \citenamefont
  {C\'eleri}, \citenamefont {Goold}, \citenamefont {Vinjanampathy},\ and\
  \citenamefont {Modi}}]{Campaioli2017}%
  \BibitemOpen
  \bibfield  {author} {\bibinfo {author} {\bibfnamefont {F.}~\bibnamefont
  {Campaioli}}, \bibinfo {author} {\bibfnamefont {F.~A.}\ \bibnamefont
  {Pollock}}, \bibinfo {author} {\bibfnamefont {F.~C.}\ \bibnamefont {Binder}},
  \bibinfo {author} {\bibfnamefont {L.}~\bibnamefont {C\'eleri}}, \bibinfo
  {author} {\bibfnamefont {J.}~\bibnamefont {Goold}}, \bibinfo {author}
  {\bibfnamefont {S.}~\bibnamefont {Vinjanampathy}},\ and\ \bibinfo {author}
  {\bibfnamefont {K.}~\bibnamefont {Modi}},\ }\bibfield  {title} {\bibinfo
  {title} {Enhancing the charging power of quantum batteries},\ }\href
  {https://doi.org/10.1103/PhysRevLett.118.150601} {\bibfield  {journal}
  {\bibinfo  {journal} {Phys. Rev. Lett.}\ }\textbf {\bibinfo {volume} {118}},\
  \bibinfo {pages} {150601} (\bibinfo {year} {2017})}\BibitemShut {NoStop}%
\bibitem [{\citenamefont {Ferraro}\ \emph {et~al.}(2018)\citenamefont
  {Ferraro}, \citenamefont {Campisi}, \citenamefont {Andolina}, \citenamefont
  {Pellegrini},\ and\ \citenamefont {Polini}}]{Ferraro2018}%
  \BibitemOpen
  \bibfield  {author} {\bibinfo {author} {\bibfnamefont {D.}~\bibnamefont
  {Ferraro}}, \bibinfo {author} {\bibfnamefont {M.}~\bibnamefont {Campisi}},
  \bibinfo {author} {\bibfnamefont {G.~M.}\ \bibnamefont {Andolina}}, \bibinfo
  {author} {\bibfnamefont {V.}~\bibnamefont {Pellegrini}},\ and\ \bibinfo
  {author} {\bibfnamefont {M.}~\bibnamefont {Polini}},\ }\bibfield  {title}
  {\bibinfo {title} {High-power collective charging of a solid-state quantum
  battery},\ }\href {https://doi.org/10.1103/PhysRevLett.120.117702} {\bibfield
   {journal} {\bibinfo  {journal} {Phys. Rev. Lett.}\ }\textbf {\bibinfo
  {volume} {120}},\ \bibinfo {pages} {117702} (\bibinfo {year}
  {2018})}\BibitemShut {NoStop}%
\bibitem [{\citenamefont {Andolina}\ \emph {et~al.}(2019)\citenamefont
  {Andolina}, \citenamefont {Keck}, \citenamefont {Mari}, \citenamefont
  {Campisi}, \citenamefont {Giovannetti},\ and\ \citenamefont
  {Polini}}]{Andolina_prl_2019}%
  \BibitemOpen
  \bibfield  {author} {\bibinfo {author} {\bibfnamefont {G.~M.}\ \bibnamefont
  {Andolina}}, \bibinfo {author} {\bibfnamefont {M.}~\bibnamefont {Keck}},
  \bibinfo {author} {\bibfnamefont {A.}~\bibnamefont {Mari}}, \bibinfo {author}
  {\bibfnamefont {M.}~\bibnamefont {Campisi}}, \bibinfo {author} {\bibfnamefont
  {V.}~\bibnamefont {Giovannetti}},\ and\ \bibinfo {author} {\bibfnamefont
  {M.}~\bibnamefont {Polini}},\ }\bibfield  {title} {\bibinfo {title}
  {Extractable work, the role of correlations, and asymptotic freedom in
  quantum batteries},\ }\href {https://doi.org/10.1103/PhysRevLett.122.047702}
  {\bibfield  {journal} {\bibinfo  {journal} {Phys. Rev. Lett.}\ }\textbf
  {\bibinfo {volume} {122}},\ \bibinfo {pages} {047702} (\bibinfo {year}
  {2019})}\BibitemShut {NoStop}%
\bibitem [{\citenamefont {Caravelli}\ \emph {et~al.}(2020)\citenamefont
  {Caravelli}, \citenamefont {Coulter-De~Wit}, \citenamefont
  {Garc\'{\i}a-Pintos},\ and\ \citenamefont {Hamma}}]{caravelli_prr_2020}%
  \BibitemOpen
  \bibfield  {author} {\bibinfo {author} {\bibfnamefont {F.}~\bibnamefont
  {Caravelli}}, \bibinfo {author} {\bibfnamefont {G.}~\bibnamefont
  {Coulter-De~Wit}}, \bibinfo {author} {\bibfnamefont {L.~P.}\ \bibnamefont
  {Garc\'{\i}a-Pintos}},\ and\ \bibinfo {author} {\bibfnamefont
  {A.}~\bibnamefont {Hamma}},\ }\bibfield  {title} {\bibinfo {title} {Random
  quantum batteries},\ }\href
  {https://doi.org/10.1103/PhysRevResearch.2.023095} {\bibfield  {journal}
  {\bibinfo  {journal} {Phys. Rev. Res.}\ }\textbf {\bibinfo {volume} {2}},\
  \bibinfo {pages} {023095} (\bibinfo {year} {2020})}\BibitemShut {NoStop}%
\bibitem [{\citenamefont {Shi}\ \emph {et~al.}(2022)\citenamefont {Shi},
  \citenamefont {Ding}, \citenamefont {Wan}, \citenamefont {Wang},\ and\
  \citenamefont {Yang}}]{Shi2022}%
  \BibitemOpen
  \bibfield  {author} {\bibinfo {author} {\bibfnamefont {H.-L.}\ \bibnamefont
  {Shi}}, \bibinfo {author} {\bibfnamefont {S.}~\bibnamefont {Ding}}, \bibinfo
  {author} {\bibfnamefont {Q.-K.}\ \bibnamefont {Wan}}, \bibinfo {author}
  {\bibfnamefont {X.-H.}\ \bibnamefont {Wang}},\ and\ \bibinfo {author}
  {\bibfnamefont {W.-L.}\ \bibnamefont {Yang}},\ }\bibfield  {title} {\bibinfo
  {title} {Entanglement, coherence, and extractable work in quantum
  batteries},\ }\href {https://doi.org/10.1103/PhysRevLett.129.130602}
  {\bibfield  {journal} {\bibinfo  {journal} {Phys. Rev. Lett.}\ }\textbf
  {\bibinfo {volume} {129}},\ \bibinfo {pages} {130602} (\bibinfo {year}
  {2022})}\BibitemShut {NoStop}%
\bibitem [{\citenamefont {Konar}\ \emph {et~al.}(2022)\citenamefont {Konar},
  \citenamefont {Lakkaraju}, \citenamefont {Ghosh},\ and\ \citenamefont
  {Sen(De)}}]{Konar2022}%
  \BibitemOpen
  \bibfield  {author} {\bibinfo {author} {\bibfnamefont {T.~K.}\ \bibnamefont
  {Konar}}, \bibinfo {author} {\bibfnamefont {L.~G.~C.}\ \bibnamefont
  {Lakkaraju}}, \bibinfo {author} {\bibfnamefont {S.}~\bibnamefont {Ghosh}},\
  and\ \bibinfo {author} {\bibfnamefont {A.}~\bibnamefont {Sen(De)}},\
  }\bibfield  {title} {\bibinfo {title} {Quantum battery with ultracold atoms:
  Bosons versus fermions},\ }\href
  {https://doi.org/10.1103/PhysRevA.106.022618} {\bibfield  {journal} {\bibinfo
   {journal} {Phys. Rev. A}\ }\textbf {\bibinfo {volume} {106}},\ \bibinfo
  {pages} {022618} (\bibinfo {year} {2022})}\BibitemShut {NoStop}%
\bibitem [{\citenamefont {Juli\`a-Farr\'e}\ \emph {et~al.}(2020)\citenamefont
  {Juli\`a-Farr\'e}, \citenamefont {Salamon}, \citenamefont {Riera},
  \citenamefont {Bera},\ and\ \citenamefont {Lewenstein}}]{Farre2020}%
  \BibitemOpen
  \bibfield  {author} {\bibinfo {author} {\bibfnamefont {S.}~\bibnamefont
  {Juli\`a-Farr\'e}}, \bibinfo {author} {\bibfnamefont {T.}~\bibnamefont
  {Salamon}}, \bibinfo {author} {\bibfnamefont {A.}~\bibnamefont {Riera}},
  \bibinfo {author} {\bibfnamefont {M.~N.}\ \bibnamefont {Bera}},\ and\
  \bibinfo {author} {\bibfnamefont {M.}~\bibnamefont {Lewenstein}},\ }\bibfield
   {title} {\bibinfo {title} {Bounds on the capacity and power of quantum
  batteries},\ }\href {https://doi.org/10.1103/PhysRevResearch.2.023113}
  {\bibfield  {journal} {\bibinfo  {journal} {Phys. Rev. Res.}\ }\textbf
  {\bibinfo {volume} {2}},\ \bibinfo {pages} {023113} (\bibinfo {year}
  {2020})}\BibitemShut {NoStop}%
\bibitem [{\citenamefont {Yang}\ \emph {et~al.}(2023)\citenamefont {Yang},
  \citenamefont {Yang}, \citenamefont {Alimuddin}, \citenamefont {Salvia},
  \citenamefont {Fei}, \citenamefont {Zhao}, \citenamefont {Nimmrichter},\ and\
  \citenamefont {Luo}}]{Yang2023}%
  \BibitemOpen
  \bibfield  {author} {\bibinfo {author} {\bibfnamefont {X.}~\bibnamefont
  {Yang}}, \bibinfo {author} {\bibfnamefont {Y.-H.}\ \bibnamefont {Yang}},
  \bibinfo {author} {\bibfnamefont {M.}~\bibnamefont {Alimuddin}}, \bibinfo
  {author} {\bibfnamefont {R.}~\bibnamefont {Salvia}}, \bibinfo {author}
  {\bibfnamefont {S.-M.}\ \bibnamefont {Fei}}, \bibinfo {author} {\bibfnamefont
  {L.-M.}\ \bibnamefont {Zhao}}, \bibinfo {author} {\bibfnamefont
  {S.}~\bibnamefont {Nimmrichter}},\ and\ \bibinfo {author} {\bibfnamefont
  {M.-X.}\ \bibnamefont {Luo}},\ }\bibfield  {title} {\bibinfo {title} {Battery
  capacity of energy-storing quantum systems},\ }\href
  {https://doi.org/10.1103/PhysRevLett.131.030402} {\bibfield  {journal}
  {\bibinfo  {journal} {Phys. Rev. Lett.}\ }\textbf {\bibinfo {volume} {131}},\
  \bibinfo {pages} {030402} (\bibinfo {year} {2023})}\BibitemShut {NoStop}%
\bibitem [{\citenamefont {Imai}\ \emph {et~al.}(2023)\citenamefont {Imai},
  \citenamefont {G\"uhne},\ and\ \citenamefont {Nimmrichter}}]{imai_pra_2023}%
  \BibitemOpen
  \bibfield  {author} {\bibinfo {author} {\bibfnamefont {S.}~\bibnamefont
  {Imai}}, \bibinfo {author} {\bibfnamefont {O.}~\bibnamefont {G\"uhne}},\ and\
  \bibinfo {author} {\bibfnamefont {S.}~\bibnamefont {Nimmrichter}},\
  }\bibfield  {title} {\bibinfo {title} {Work fluctuations and entanglement in
  quantum batteries},\ }\href {https://doi.org/10.1103/PhysRevA.107.022215}
  {\bibfield  {journal} {\bibinfo  {journal} {Phys. Rev. A}\ }\textbf {\bibinfo
  {volume} {107}},\ \bibinfo {pages} {022215} (\bibinfo {year}
  {2023})}\BibitemShut {NoStop}%
\bibitem [{\citenamefont {Bhanja}\ \emph {et~al.}(2024)\citenamefont {Bhanja},
  \citenamefont {Tiwari},\ and\ \citenamefont {Banerjee}}]{Bhanja_pra_2024}%
  \BibitemOpen
  \bibfield  {author} {\bibinfo {author} {\bibfnamefont {G.}~\bibnamefont
  {Bhanja}}, \bibinfo {author} {\bibfnamefont {D.}~\bibnamefont {Tiwari}},\
  and\ \bibinfo {author} {\bibfnamefont {S.}~\bibnamefont {Banerjee}},\
  }\bibfield  {title} {\bibinfo {title} {Impact of non-markovian quantum
  brownian motion on quantum batteries},\ }\href
  {https://doi.org/10.1103/PhysRevA.109.012224} {\bibfield  {journal} {\bibinfo
   {journal} {Phys. Rev. A}\ }\textbf {\bibinfo {volume} {109}},\ \bibinfo
  {pages} {012224} (\bibinfo {year} {2024})}\BibitemShut {NoStop}%
\bibitem [{\citenamefont {Vigneshwar}\ and\ \citenamefont
  {Sankaranarayanan}(2025)}]{vigneshwar2025}%
  \BibitemOpen
  \bibfield  {author} {\bibinfo {author} {\bibfnamefont {B.}~\bibnamefont
  {Vigneshwar}}\ and\ \bibinfo {author} {\bibfnamefont {R.}~\bibnamefont
  {Sankaranarayanan}},\ }\href {https://arxiv.org/abs/2512.14497} {\bibinfo
  {title} {Nonlocal contributions to ergotropy: A thermodynamic perspective}}
  (\bibinfo {year} {2025}),\ \Eprint {https://arxiv.org/abs/2512.14497}
  {arXiv:2512.14497 [quant-ph]} \BibitemShut {NoStop}%
\bibitem [{\citenamefont {Le}\ \emph {et~al.}(2018)\citenamefont {Le},
  \citenamefont {Levinsen}, \citenamefont {Modi}, \citenamefont {Parish},\ and\
  \citenamefont {Pollock}}]{Le_pra_2018}%
  \BibitemOpen
  \bibfield  {author} {\bibinfo {author} {\bibfnamefont {T.~P.}\ \bibnamefont
  {Le}}, \bibinfo {author} {\bibfnamefont {J.}~\bibnamefont {Levinsen}},
  \bibinfo {author} {\bibfnamefont {K.}~\bibnamefont {Modi}}, \bibinfo {author}
  {\bibfnamefont {M.~M.}\ \bibnamefont {Parish}},\ and\ \bibinfo {author}
  {\bibfnamefont {F.~A.}\ \bibnamefont {Pollock}},\ }\bibfield  {title}
  {\bibinfo {title} {Spin-chain model of a many-body quantum battery},\ }\href
  {https://doi.org/10.1103/PhysRevA.97.022106} {\bibfield  {journal} {\bibinfo
  {journal} {Phys. Rev. A}\ }\textbf {\bibinfo {volume} {97}},\ \bibinfo
  {pages} {022106} (\bibinfo {year} {2018})}\BibitemShut {NoStop}%
\bibitem [{\citenamefont {Rossini}\ \emph {et~al.}(2019)\citenamefont
  {Rossini}, \citenamefont {Andolina},\ and\ \citenamefont
  {Polini}}]{rossini_prb_2019}%
  \BibitemOpen
  \bibfield  {author} {\bibinfo {author} {\bibfnamefont {D.}~\bibnamefont
  {Rossini}}, \bibinfo {author} {\bibfnamefont {G.~M.}\ \bibnamefont
  {Andolina}},\ and\ \bibinfo {author} {\bibfnamefont {M.}~\bibnamefont
  {Polini}},\ }\bibfield  {title} {\bibinfo {title} {Many-body localized
  quantum batteries},\ }\href {https://doi.org/10.1103/PhysRevB.100.115142}
  {\bibfield  {journal} {\bibinfo  {journal} {Phys. Rev. B}\ }\textbf {\bibinfo
  {volume} {100}},\ \bibinfo {pages} {115142} (\bibinfo {year}
  {2019})}\BibitemShut {NoStop}%
\bibitem [{\citenamefont {Ghosh}\ \emph {et~al.}(2020)\citenamefont {Ghosh},
  \citenamefont {Chanda},\ and\ \citenamefont {Sen(De)}}]{Ghosh2020}%
  \BibitemOpen
  \bibfield  {author} {\bibinfo {author} {\bibfnamefont {S.}~\bibnamefont
  {Ghosh}}, \bibinfo {author} {\bibfnamefont {T.}~\bibnamefont {Chanda}},\ and\
  \bibinfo {author} {\bibfnamefont {A.}~\bibnamefont {Sen(De)}},\ }\bibfield
  {title} {\bibinfo {title} {Enhancement in the performance of a quantum
  battery by ordered and disordered interactions},\ }\href
  {https://doi.org/10.1103/PhysRevA.101.032115} {\bibfield  {journal} {\bibinfo
   {journal} {Phys. Rev. A}\ }\textbf {\bibinfo {volume} {101}},\ \bibinfo
  {pages} {032115} (\bibinfo {year} {2020})}\BibitemShut {NoStop}%
\bibitem [{\citenamefont {Mondal}\ and\ \citenamefont
  {Bhattacharjee}(2022)}]{mondal_pre_2022}%
  \BibitemOpen
  \bibfield  {author} {\bibinfo {author} {\bibfnamefont {S.}~\bibnamefont
  {Mondal}}\ and\ \bibinfo {author} {\bibfnamefont {S.}~\bibnamefont
  {Bhattacharjee}},\ }\bibfield  {title} {\bibinfo {title} {Periodically driven
  many-body quantum battery},\ }\href
  {https://doi.org/10.1103/PhysRevE.105.044125} {\bibfield  {journal} {\bibinfo
   {journal} {Phys. Rev. E}\ }\textbf {\bibinfo {volume} {105}},\ \bibinfo
  {pages} {044125} (\bibinfo {year} {2022})}\BibitemShut {NoStop}%
\bibitem [{\citenamefont {Guo}\ \emph {et~al.}(2024)\citenamefont {Guo},
  \citenamefont {Yang},\ and\ \citenamefont {Dou}}]{guo_pra_2024}%
  \BibitemOpen
  \bibfield  {author} {\bibinfo {author} {\bibfnamefont {W.-X.}\ \bibnamefont
  {Guo}}, \bibinfo {author} {\bibfnamefont {F.-M.}\ \bibnamefont {Yang}},\ and\
  \bibinfo {author} {\bibfnamefont {F.-Q.}\ \bibnamefont {Dou}},\ }\bibfield
  {title} {\bibinfo {title} {Analytically solvable many-body rosen-zener
  quantum battery},\ }\href {https://doi.org/10.1103/PhysRevA.109.032201}
  {\bibfield  {journal} {\bibinfo  {journal} {Phys. Rev. A}\ }\textbf {\bibinfo
  {volume} {109}},\ \bibinfo {pages} {032201} (\bibinfo {year}
  {2024})}\BibitemShut {NoStop}%
\bibitem [{\citenamefont {Zhang}\ \emph {et~al.}(2024)\citenamefont {Zhang},
  \citenamefont {Song},\ and\ \citenamefont {Wang}}]{Zhang2024}%
  \BibitemOpen
  \bibfield  {author} {\bibinfo {author} {\bibfnamefont {X.}~\bibnamefont
  {Zhang}}, \bibinfo {author} {\bibfnamefont {X.}~\bibnamefont {Song}},\ and\
  \bibinfo {author} {\bibfnamefont {D.}~\bibnamefont {Wang}},\ }\bibfield
  {title} {\bibinfo {title} {Quantum battery in the heisenberg spin chain
  models with dzyaloshinskii‐moriya interaction},\ }\bibfield  {journal}
  {\bibinfo  {journal} {Advanced Quantum Technologies}\ }\textbf {\bibinfo
  {volume} {7}},\ \href {https://doi.org/10.1002/qute.202400114}
  {10.1002/qute.202400114} (\bibinfo {year} {2024})\BibitemShut {NoStop}%
\bibitem [{\citenamefont {Gyhm}\ and\ \citenamefont
  {Fischer}(2024)}]{Gyhm2024Jan}%
  \BibitemOpen
  \bibfield  {author} {\bibinfo {author} {\bibfnamefont {J.-Y.}\ \bibnamefont
  {Gyhm}}\ and\ \bibinfo {author} {\bibfnamefont {U.~R.}\ \bibnamefont
  {Fischer}},\ }\bibfield  {title} {\bibinfo {title} {{Beneficial and
  detrimental entanglement for quantum battery charging}},\ }\href
  {https://doi.org/10.1116/5.0184903} {\bibfield  {journal} {\bibinfo
  {journal} {AVS Quantum Sci.}\ }\textbf {\bibinfo {volume} {6}},\ \bibinfo
  {pages} {012001} (\bibinfo {year} {2024})}\BibitemShut {NoStop}%
\bibitem [{\citenamefont {Puri}\ \emph {et~al.}(2025)\citenamefont {Puri},
  \citenamefont {Konar}, \citenamefont {Lakkaraju},\ and\ \citenamefont
  {De}}]{puri2025}%
  \BibitemOpen
  \bibfield  {author} {\bibinfo {author} {\bibfnamefont {S.}~\bibnamefont
  {Puri}}, \bibinfo {author} {\bibfnamefont {T.~K.}\ \bibnamefont {Konar}},
  \bibinfo {author} {\bibfnamefont {L.~G.~C.}\ \bibnamefont {Lakkaraju}},\ and\
  \bibinfo {author} {\bibfnamefont {A.~S.}\ \bibnamefont {De}},\ }\href
  {https://arxiv.org/abs/2412.00921} {\bibinfo {title} {Floquet driven
  long-range interactions induce super-extensive scaling in quantum batteries}}
  (\bibinfo {year} {2025}),\ \Eprint {https://arxiv.org/abs/2412.00921}
  {arXiv:2412.00921 [quant-ph]} \BibitemShut {NoStop}%
\bibitem [{\citenamefont {Sahoo}\ and\ \citenamefont
  {Rakshit}(2025)}]{sahoo2025}%
  \BibitemOpen
  \bibfield  {author} {\bibinfo {author} {\bibfnamefont {A.}~\bibnamefont
  {Sahoo}}\ and\ \bibinfo {author} {\bibfnamefont {D.}~\bibnamefont
  {Rakshit}},\ }\href {https://arxiv.org/abs/2508.14847} {\bibinfo {title}
  {Power-law interactions stabilize time crystals realizing quantum energy
  storage and sensing}} (\bibinfo {year} {2025}),\ \Eprint
  {https://arxiv.org/abs/2508.14847} {arXiv:2508.14847 [quant-ph]} \BibitemShut
  {NoStop}%
\bibitem [{\citenamefont {Bhattacharya}\ \emph {et~al.}(2026)\citenamefont
  {Bhattacharya}, \citenamefont {Sabale},\ and\ \citenamefont
  {Kumar}}]{Bhattacharya2026}%
  \BibitemOpen
  \bibfield  {author} {\bibinfo {author} {\bibfnamefont {S.}~\bibnamefont
  {Bhattacharya}}, \bibinfo {author} {\bibfnamefont {V.~B.}\ \bibnamefont
  {Sabale}},\ and\ \bibinfo {author} {\bibfnamefont {A.}~\bibnamefont
  {Kumar}},\ }\bibfield  {title} {\bibinfo {title} {Heisenberg spin networks
  for realizing quantum battery with the aid of dzyaloshinskii–moriya
  interaction},\ }\href {https://doi.org/10.1088/1367-2630/ae33e4} {\bibfield
  {journal} {\bibinfo  {journal} {New Journal of Physics}\ }\textbf {\bibinfo
  {volume} {28}},\ \bibinfo {pages} {014508} (\bibinfo {year}
  {2026})}\BibitemShut {NoStop}%
\bibitem [{\citenamefont {Lu}\ \emph {et~al.}(2021)\citenamefont {Lu},
  \citenamefont {Chen}, \citenamefont {Kuang},\ and\ \citenamefont
  {Wang}}]{lu_pra_2021}%
  \BibitemOpen
  \bibfield  {author} {\bibinfo {author} {\bibfnamefont {W.}~\bibnamefont
  {Lu}}, \bibinfo {author} {\bibfnamefont {J.}~\bibnamefont {Chen}}, \bibinfo
  {author} {\bibfnamefont {L.-M.}\ \bibnamefont {Kuang}},\ and\ \bibinfo
  {author} {\bibfnamefont {X.}~\bibnamefont {Wang}},\ }\bibfield  {title}
  {\bibinfo {title} {Optimal state for a tavis-cummings quantum battery via the
  bethe ansatz method},\ }\href {https://doi.org/10.1103/PhysRevA.104.043706}
  {\bibfield  {journal} {\bibinfo  {journal} {Phys. Rev. A}\ }\textbf {\bibinfo
  {volume} {104}},\ \bibinfo {pages} {043706} (\bibinfo {year}
  {2021})}\BibitemShut {NoStop}%
\bibitem [{\citenamefont {Dou}\ \emph {et~al.}(2022)\citenamefont {Dou},
  \citenamefont {Zhou},\ and\ \citenamefont {Sun}}]{dou_pra_2022}%
  \BibitemOpen
  \bibfield  {author} {\bibinfo {author} {\bibfnamefont {F.-Q.}\ \bibnamefont
  {Dou}}, \bibinfo {author} {\bibfnamefont {H.}~\bibnamefont {Zhou}},\ and\
  \bibinfo {author} {\bibfnamefont {J.-A.}\ \bibnamefont {Sun}},\ }\bibfield
  {title} {\bibinfo {title} {Cavity heisenberg-spin-chain quantum battery},\
  }\href {https://doi.org/10.1103/PhysRevA.106.032212} {\bibfield  {journal}
  {\bibinfo  {journal} {Phys. Rev. A}\ }\textbf {\bibinfo {volume} {106}},\
  \bibinfo {pages} {032212} (\bibinfo {year} {2022})}\BibitemShut {NoStop}%
\bibitem [{\citenamefont {Zhao}\ \emph {et~al.}(2025)\citenamefont {Zhao},
  \citenamefont {Zhao},\ and\ \citenamefont {Zhuang}}]{zhao_pra_2025}%
  \BibitemOpen
  \bibfield  {author} {\bibinfo {author} {\bibfnamefont {S.-C.}\ \bibnamefont
  {Zhao}}, \bibinfo {author} {\bibfnamefont {Z.-R.}\ \bibnamefont {Zhao}},\
  and\ \bibinfo {author} {\bibfnamefont {N.-Y.}\ \bibnamefont {Zhuang}},\
  }\bibfield  {title} {\bibinfo {title} {Non-markovian $n$-spin chain quantum
  battery in thermal charging process},\ }\href
  {https://doi.org/10.1103/xqtv-qbyk} {\bibfield  {journal} {\bibinfo
  {journal} {Phys. Rev. E}\ }\textbf {\bibinfo {volume} {112}},\ \bibinfo
  {pages} {024129} (\bibinfo {year} {2025})}\BibitemShut {NoStop}%
\bibitem [{\citenamefont {Pushpan}\ and\ \citenamefont
  {Pal}(2025)}]{pushpan2025}%
  \BibitemOpen
  \bibfield  {author} {\bibinfo {author} {\bibfnamefont {C.~B.}\ \bibnamefont
  {Pushpan}}\ and\ \bibinfo {author} {\bibfnamefont {A.~K.}\ \bibnamefont
  {Pal}},\ }\href {https://arxiv.org/abs/2511.07243} {\bibinfo {title}
  {Bridging the daemonic gap en route to charge multi-mode batteries via a
  single auxiliary}} (\bibinfo {year} {2025}),\ \Eprint
  {https://arxiv.org/abs/2511.07243} {arXiv:2511.07243 [quant-ph]} \BibitemShut
  {NoStop}%
\bibitem [{\citenamefont {Ghosh}\ \emph {et~al.}(2021)\citenamefont {Ghosh},
  \citenamefont {Chanda}, \citenamefont {Mal},\ and\ \citenamefont
  {Sen(De)}}]{Ghosh2021}%
  \BibitemOpen
  \bibfield  {author} {\bibinfo {author} {\bibfnamefont {S.}~\bibnamefont
  {Ghosh}}, \bibinfo {author} {\bibfnamefont {T.}~\bibnamefont {Chanda}},
  \bibinfo {author} {\bibfnamefont {S.}~\bibnamefont {Mal}},\ and\ \bibinfo
  {author} {\bibfnamefont {A.}~\bibnamefont {Sen(De)}},\ }\bibfield  {title}
  {\bibinfo {title} {Fast charging of a quantum battery assisted by noise},\
  }\href {https://doi.org/10.1103/PhysRevA.104.032207} {\bibfield  {journal}
  {\bibinfo  {journal} {Phys. Rev. A}\ }\textbf {\bibinfo {volume} {104}},\
  \bibinfo {pages} {032207} (\bibinfo {year} {2021})}\BibitemShut {NoStop}%
\bibitem [{\citenamefont {Zakavati}\ \emph {et~al.}(2021)\citenamefont
  {Zakavati}, \citenamefont {Tabesh},\ and\ \citenamefont
  {Salimi}}]{Zakavati2021}%
  \BibitemOpen
  \bibfield  {author} {\bibinfo {author} {\bibfnamefont {S.}~\bibnamefont
  {Zakavati}}, \bibinfo {author} {\bibfnamefont {F.~T.}\ \bibnamefont
  {Tabesh}},\ and\ \bibinfo {author} {\bibfnamefont {S.}~\bibnamefont
  {Salimi}},\ }\bibfield  {title} {\bibinfo {title} {Bounds on charging power
  of open quantum batteries},\ }\href
  {https://doi.org/10.1103/PhysRevE.104.054117} {\bibfield  {journal} {\bibinfo
   {journal} {Phys. Rev. E}\ }\textbf {\bibinfo {volume} {104}},\ \bibinfo
  {pages} {054117} (\bibinfo {year} {2021})}\BibitemShut {NoStop}%
\bibitem [{\citenamefont {Arjmandi}\ \emph {et~al.}(2022)\citenamefont
  {Arjmandi}, \citenamefont {Mohammadi},\ and\ \citenamefont
  {Santos}}]{Arjmandi2022}%
  \BibitemOpen
  \bibfield  {author} {\bibinfo {author} {\bibfnamefont {M.~B.}\ \bibnamefont
  {Arjmandi}}, \bibinfo {author} {\bibfnamefont {H.}~\bibnamefont
  {Mohammadi}},\ and\ \bibinfo {author} {\bibfnamefont {A.~C.}\ \bibnamefont
  {Santos}},\ }\bibfield  {title} {\bibinfo {title} {Enhancing self-discharging
  process with disordered quantum batteries},\ }\href
  {https://doi.org/10.1103/PhysRevE.105.054115} {\bibfield  {journal} {\bibinfo
   {journal} {Phys. Rev. E}\ }\textbf {\bibinfo {volume} {105}},\ \bibinfo
  {pages} {054115} (\bibinfo {year} {2022})}\BibitemShut {NoStop}%
\bibitem [{\citenamefont {Liu}\ \emph {et~al.}(2024)\citenamefont {Liu},
  \citenamefont {Wang}, \citenamefont {Fan}, \citenamefont {Wu},\ and\
  \citenamefont {Liu}}]{Liu2024}%
  \BibitemOpen
  \bibfield  {author} {\bibinfo {author} {\bibfnamefont {S.-Q.}\ \bibnamefont
  {Liu}}, \bibinfo {author} {\bibfnamefont {L.}~\bibnamefont {Wang}}, \bibinfo
  {author} {\bibfnamefont {H.}~\bibnamefont {Fan}}, \bibinfo {author}
  {\bibfnamefont {F.-L.}\ \bibnamefont {Wu}},\ and\ \bibinfo {author}
  {\bibfnamefont {S.-Y.}\ \bibnamefont {Liu}},\ }\bibfield  {title} {\bibinfo
  {title} {Better performance of quantum batteries in different environments
  compared to closed batteries},\ }\href
  {https://doi.org/10.1103/PhysRevA.109.042411} {\bibfield  {journal} {\bibinfo
   {journal} {Phys. Rev. A}\ }\textbf {\bibinfo {volume} {109}},\ \bibinfo
  {pages} {042411} (\bibinfo {year} {2024})}\BibitemShut {NoStop}%
\bibitem [{\citenamefont {Tirone}\ \emph {et~al.}(2023)\citenamefont {Tirone},
  \citenamefont {Salvia}, \citenamefont {Chessa},\ and\ \citenamefont
  {Giovannetti}}]{Tirone2023}%
  \BibitemOpen
  \bibfield  {author} {\bibinfo {author} {\bibfnamefont {S.}~\bibnamefont
  {Tirone}}, \bibinfo {author} {\bibfnamefont {R.}~\bibnamefont {Salvia}},
  \bibinfo {author} {\bibfnamefont {S.}~\bibnamefont {Chessa}},\ and\ \bibinfo
  {author} {\bibfnamefont {V.}~\bibnamefont {Giovannetti}},\ }\bibfield
  {title} {\bibinfo {title} {Work extraction processes from noisy quantum
  batteries: The role of nonlocal resources},\ }\href
  {https://doi.org/10.1103/PhysRevLett.131.060402} {\bibfield  {journal}
  {\bibinfo  {journal} {Phys. Rev. Lett.}\ }\textbf {\bibinfo {volume} {131}},\
  \bibinfo {pages} {060402} (\bibinfo {year} {2023})}\BibitemShut {NoStop}%
\bibitem [{\citenamefont {Sarkar}\ \emph {et~al.}(2025)\citenamefont {Sarkar},
  \citenamefont {Chaki}, \citenamefont {Ghosh},\ and\ \citenamefont
  {Sen}}]{sarkar2025}%
  \BibitemOpen
  \bibfield  {author} {\bibinfo {author} {\bibfnamefont {A.}~\bibnamefont
  {Sarkar}}, \bibinfo {author} {\bibfnamefont {P.}~\bibnamefont {Chaki}},
  \bibinfo {author} {\bibfnamefont {P.}~\bibnamefont {Ghosh}},\ and\ \bibinfo
  {author} {\bibfnamefont {U.}~\bibnamefont {Sen}},\ }\href
  {https://arxiv.org/abs/2505.16851} {\bibinfo {title} {Fluctuation in energy
  extraction from quantum batteries: How open should the system be to control
  it?}} (\bibinfo {year} {2025}),\ \Eprint {https://arxiv.org/abs/2505.16851}
  {arXiv:2505.16851 [quant-ph]} \BibitemShut {NoStop}%
\bibitem [{\citenamefont {Kamin}\ \emph {et~al.}(2020)\citenamefont {Kamin},
  \citenamefont {Tabesh}, \citenamefont {Salimi}, \citenamefont {Kheirandish},\
  and\ \citenamefont {Santos}}]{Kamin2020}%
  \BibitemOpen
  \bibfield  {author} {\bibinfo {author} {\bibfnamefont {F.~H.}\ \bibnamefont
  {Kamin}}, \bibinfo {author} {\bibfnamefont {F.~T.}\ \bibnamefont {Tabesh}},
  \bibinfo {author} {\bibfnamefont {S.}~\bibnamefont {Salimi}}, \bibinfo
  {author} {\bibfnamefont {F.}~\bibnamefont {Kheirandish}},\ and\ \bibinfo
  {author} {\bibfnamefont {A.~C.}\ \bibnamefont {Santos}},\ }\bibfield  {title}
  {\bibinfo {title} {Non-markovian effects on charging and self-discharging
  process of quantum batteries},\ }\href
  {https://doi.org/10.1088/1367-2630/ab9ee2} {\bibfield  {journal} {\bibinfo
  {journal} {New Journal of Physics}\ }\textbf {\bibinfo {volume} {22}},\
  \bibinfo {pages} {083007} (\bibinfo {year} {2020})}\BibitemShut {NoStop}%
\bibitem [{\citenamefont {Santos}(2021)}]{Santos2021}%
  \BibitemOpen
  \bibfield  {author} {\bibinfo {author} {\bibfnamefont {A.~C.}\ \bibnamefont
  {Santos}},\ }\bibfield  {title} {\bibinfo {title} {Quantum advantage of
  two-level batteries in the self-discharging process},\ }\href
  {https://doi.org/10.1103/PhysRevE.103.042118} {\bibfield  {journal} {\bibinfo
   {journal} {Phys. Rev. E}\ }\textbf {\bibinfo {volume} {103}},\ \bibinfo
  {pages} {042118} (\bibinfo {year} {2021})}\BibitemShut {NoStop}%
\bibitem [{\citenamefont {Xu}\ \emph {et~al.}(2024)\citenamefont {Xu},
  \citenamefont {Li}, \citenamefont {Zhu},\ and\ \citenamefont {Liu}}]{Xu2024}%
  \BibitemOpen
  \bibfield  {author} {\bibinfo {author} {\bibfnamefont {K.}~\bibnamefont
  {Xu}}, \bibinfo {author} {\bibfnamefont {H.-G.}\ \bibnamefont {Li}}, \bibinfo
  {author} {\bibfnamefont {H.-J.}\ \bibnamefont {Zhu}},\ and\ \bibinfo {author}
  {\bibfnamefont {W.-M.}\ \bibnamefont {Liu}},\ }\bibfield  {title} {\bibinfo
  {title} {Inhibiting the self-discharging process of quantum batteries in
  non-markovian noises},\ }\href {https://doi.org/10.1103/PhysRevE.109.054132}
  {\bibfield  {journal} {\bibinfo  {journal} {Phys. Rev. E}\ }\textbf {\bibinfo
  {volume} {109}},\ \bibinfo {pages} {054132} (\bibinfo {year}
  {2024})}\BibitemShut {NoStop}%
\bibitem [{\citenamefont {Chaki}\ \emph {et~al.}(2025)\citenamefont {Chaki},
  \citenamefont {Bhattacharyya}, \citenamefont {Sen},\ and\ \citenamefont
  {Sen}}]{Chaki2025}%
  \BibitemOpen
  \bibfield  {author} {\bibinfo {author} {\bibfnamefont {P.}~\bibnamefont
  {Chaki}}, \bibinfo {author} {\bibfnamefont {A.}~\bibnamefont
  {Bhattacharyya}}, \bibinfo {author} {\bibfnamefont {K.}~\bibnamefont {Sen}},\
  and\ \bibinfo {author} {\bibfnamefont {U.}~\bibnamefont {Sen}},\ }\bibfield
  {title} {\bibinfo {title} {Auxiliary-assisted energy distillation from
  quantum batteries},\ }\href {https://doi.org/10.1103/cyrc-ms34} {\bibfield
  {journal} {\bibinfo  {journal} {Phys. Rev. A}\ }\textbf {\bibinfo {volume}
  {112}},\ \bibinfo {pages} {052446} (\bibinfo {year} {2025})}\BibitemShut
  {NoStop}%
\bibitem [{\citenamefont {Morrone}\ \emph {et~al.}(2023)\citenamefont
  {Morrone}, \citenamefont {Rossi},\ and\ \citenamefont
  {Genoni}}]{Morrone2023}%
  \BibitemOpen
  \bibfield  {author} {\bibinfo {author} {\bibfnamefont {D.}~\bibnamefont
  {Morrone}}, \bibinfo {author} {\bibfnamefont {M.~A.}\ \bibnamefont {Rossi}},\
  and\ \bibinfo {author} {\bibfnamefont {M.~G.}\ \bibnamefont {Genoni}},\
  }\bibfield  {title} {\bibinfo {title} {Daemonic ergotropy in continuously
  monitored open quantum batteries},\ }\href
  {https://doi.org/10.1103/PhysRevApplied.20.044073} {\bibfield  {journal}
  {\bibinfo  {journal} {Phys. Rev. Appl.}\ }\textbf {\bibinfo {volume} {20}},\
  \bibinfo {pages} {044073} (\bibinfo {year} {2023})}\BibitemShut {NoStop}%
\bibitem [{\citenamefont {Ahmadi}\ \emph {et~al.}(2024)\citenamefont {Ahmadi},
  \citenamefont {Mazurek}, \citenamefont {Horodecki},\ and\ \citenamefont
  {Barzanjeh}}]{ahmadi_prl_2024}%
  \BibitemOpen
  \bibfield  {author} {\bibinfo {author} {\bibfnamefont {B.}~\bibnamefont
  {Ahmadi}}, \bibinfo {author} {\bibfnamefont {P.}~\bibnamefont {Mazurek}},
  \bibinfo {author} {\bibfnamefont {P.}~\bibnamefont {Horodecki}},\ and\
  \bibinfo {author} {\bibfnamefont {S.}~\bibnamefont {Barzanjeh}},\ }\bibfield
  {title} {\bibinfo {title} {Nonreciprocal quantum batteries},\ }\href
  {https://doi.org/10.1103/PhysRevLett.132.210402} {\bibfield  {journal}
  {\bibinfo  {journal} {Phys. Rev. Lett.}\ }\textbf {\bibinfo {volume} {132}},\
  \bibinfo {pages} {210402} (\bibinfo {year} {2024})}\BibitemShut {NoStop}%
\bibitem [{\citenamefont {Medina}\ \emph {et~al.}(2025)\citenamefont {Medina},
  \citenamefont {Culhane}, \citenamefont {Binder}, \citenamefont {Landi},\ and\
  \citenamefont {Goold}}]{Medina2024}%
  \BibitemOpen
  \bibfield  {author} {\bibinfo {author} {\bibfnamefont {I.}~\bibnamefont
  {Medina}}, \bibinfo {author} {\bibfnamefont {O.}~\bibnamefont {Culhane}},
  \bibinfo {author} {\bibfnamefont {F.~C.}\ \bibnamefont {Binder}}, \bibinfo
  {author} {\bibfnamefont {G.~T.}\ \bibnamefont {Landi}},\ and\ \bibinfo
  {author} {\bibfnamefont {J.}~\bibnamefont {Goold}},\ }\bibfield  {title}
  {\bibinfo {title} {Anomalous discharging of quantum batteries: The ergotropic
  mpemba effect},\ }\href {https://doi.org/10.1103/PhysRevLett.134.220402}
  {\bibfield  {journal} {\bibinfo  {journal} {Phys. Rev. Lett.}\ }\textbf
  {\bibinfo {volume} {134}},\ \bibinfo {pages} {220402} (\bibinfo {year}
  {2025})}\BibitemShut {NoStop}%
\bibitem [{\citenamefont {Hadipour}\ and\ \citenamefont
  {Haseli}(2025)}]{Maryam2025}%
  \BibitemOpen
  \bibfield  {author} {\bibinfo {author} {\bibfnamefont {M.}~\bibnamefont
  {Hadipour}}\ and\ \bibinfo {author} {\bibfnamefont {S.}~\bibnamefont
  {Haseli}},\ }\href {https://arxiv.org/abs/2502.05508} {\bibinfo {title}
  {Nonequilibrium quantum batteries: Amplified work extraction through thermal
  bath modulation}} (\bibinfo {year} {2025}),\ \Eprint
  {https://arxiv.org/abs/2502.05508} {arXiv:2502.05508 [quant-ph]} \BibitemShut
  {NoStop}%
\bibitem [{\citenamefont {Lu}\ \emph {et~al.}(2025)\citenamefont {Lu},
  \citenamefont {Tian}, \citenamefont {L\"u},\ and\ \citenamefont
  {Shang}}]{topological_quantumbattery}%
  \BibitemOpen
  \bibfield  {author} {\bibinfo {author} {\bibfnamefont {Z.-G.}\ \bibnamefont
  {Lu}}, \bibinfo {author} {\bibfnamefont {G.}~\bibnamefont {Tian}}, \bibinfo
  {author} {\bibfnamefont {X.-Y.}\ \bibnamefont {L\"u}},\ and\ \bibinfo
  {author} {\bibfnamefont {C.}~\bibnamefont {Shang}},\ }\href
  {https://doi.org/10.1103/PhysRevLett.134.180401} {\bibinfo {title}
  {Topological quantum batteries}} (\bibinfo {year} {2025})\BibitemShut
  {NoStop}%
\bibitem [{\citenamefont {Vigneshwar}\ and\ \citenamefont
  {Sankaranarayanan}(2026)}]{Vigneshwar2026}%
  \BibitemOpen
  \bibfield  {author} {\bibinfo {author} {\bibfnamefont {B.}~\bibnamefont
  {Vigneshwar}}\ and\ \bibinfo {author} {\bibfnamefont {R.}~\bibnamefont
  {Sankaranarayanan}},\ }\bibfield  {title} {\bibinfo {title} {Noise resilience
  of spin quantum battery in the presence of dm interactions},\ }\href
  {https://doi.org/10.1088/1751-8121/ae30b8} {\bibfield  {journal} {\bibinfo
  {journal} {Journal of Physics A: Mathematical and Theoretical}\ }\textbf
  {\bibinfo {volume} {59}},\ \bibinfo {pages} {015302} (\bibinfo {year}
  {2026})}\BibitemShut {NoStop}%
\bibitem [{\citenamefont {Joshi}\ and\ \citenamefont
  {Mahesh}(2022)}]{MaheshexpNMR}%
  \BibitemOpen
  \bibfield  {author} {\bibinfo {author} {\bibfnamefont {J.}~\bibnamefont
  {Joshi}}\ and\ \bibinfo {author} {\bibfnamefont {T.~S.}\ \bibnamefont
  {Mahesh}},\ }\bibfield  {title} {\bibinfo {title} {Experimental investigation
  of a quantum battery using star-topology nmr spin systems},\ }\href
  {https://doi.org/10.1103/PhysRevA.106.042601} {\bibfield  {journal} {\bibinfo
   {journal} {Phys. Rev. A}\ }\textbf {\bibinfo {volume} {106}},\ \bibinfo
  {pages} {042601} (\bibinfo {year} {2022})}\BibitemShut {NoStop}%
\bibitem [{\citenamefont {de~Buy~Wenniger}\ \emph {et~al.}(2023)\citenamefont
  {de~Buy~Wenniger}, \citenamefont {Thomas}, \citenamefont {Maffei},
  \citenamefont {Wein}, \citenamefont {Pont}, \citenamefont {Belabas},
  \citenamefont {Prasad}, \citenamefont {Harouri}, \citenamefont {Lemaître},
  \citenamefont {Sagnes}, \citenamefont {Somaschi}, \citenamefont {Auffèves},\
  and\ \citenamefont {Senellart}}]{wennigerexpqdots}%
  \BibitemOpen
  \bibfield  {author} {\bibinfo {author} {\bibfnamefont {I.~M.}\ \bibnamefont
  {de~Buy~Wenniger}}, \bibinfo {author} {\bibfnamefont {S.~E.}\ \bibnamefont
  {Thomas}}, \bibinfo {author} {\bibfnamefont {M.}~\bibnamefont {Maffei}},
  \bibinfo {author} {\bibfnamefont {S.~C.}\ \bibnamefont {Wein}}, \bibinfo
  {author} {\bibfnamefont {M.}~\bibnamefont {Pont}}, \bibinfo {author}
  {\bibfnamefont {N.}~\bibnamefont {Belabas}}, \bibinfo {author} {\bibfnamefont
  {S.}~\bibnamefont {Prasad}}, \bibinfo {author} {\bibfnamefont
  {A.}~\bibnamefont {Harouri}}, \bibinfo {author} {\bibfnamefont
  {A.}~\bibnamefont {Lemaître}}, \bibinfo {author} {\bibfnamefont
  {I.}~\bibnamefont {Sagnes}}, \bibinfo {author} {\bibfnamefont
  {N.}~\bibnamefont {Somaschi}}, \bibinfo {author} {\bibfnamefont
  {A.}~\bibnamefont {Auffèves}},\ and\ \bibinfo {author} {\bibfnamefont
  {P.}~\bibnamefont {Senellart}},\ }\href {https://arxiv.org/abs/2202.01109}
  {\bibinfo {title} {Experimental analysis of energy transfers between a
  quantum emitter and light fields}} (\bibinfo {year} {2023}),\ \Eprint
  {https://arxiv.org/abs/2202.01109} {arXiv:2202.01109 [quant-ph]} \BibitemShut
  {NoStop}%
\bibitem [{\citenamefont {Quach}\ \emph {et~al.}(2022)\citenamefont {Quach},
  \citenamefont {McGhee}, \citenamefont {Ganzer}, \citenamefont {Rouse},
  \citenamefont {Lovett}, \citenamefont {Gauger}, \citenamefont {Keeling},
  \citenamefont {Cerullo}, \citenamefont {Lidzey},\ and\ \citenamefont
  {Virgili}}]{recentexperiment}%
  \BibitemOpen
  \bibfield  {author} {\bibinfo {author} {\bibfnamefont {J.~Q.}\ \bibnamefont
  {Quach}}, \bibinfo {author} {\bibfnamefont {K.~E.}\ \bibnamefont {McGhee}},
  \bibinfo {author} {\bibfnamefont {L.}~\bibnamefont {Ganzer}}, \bibinfo
  {author} {\bibfnamefont {D.~M.}\ \bibnamefont {Rouse}}, \bibinfo {author}
  {\bibfnamefont {B.~W.}\ \bibnamefont {Lovett}}, \bibinfo {author}
  {\bibfnamefont {E.~M.}\ \bibnamefont {Gauger}}, \bibinfo {author}
  {\bibfnamefont {J.}~\bibnamefont {Keeling}}, \bibinfo {author} {\bibfnamefont
  {G.}~\bibnamefont {Cerullo}}, \bibinfo {author} {\bibfnamefont {D.~G.}\
  \bibnamefont {Lidzey}},\ and\ \bibinfo {author} {\bibfnamefont
  {T.}~\bibnamefont {Virgili}},\ }\bibfield  {title} {\bibinfo {title}
  {Superabsorption in an organic microcavity: Toward a quantum battery},\
  }\href {https://doi.org/10.1126/sciadv.abk3160} {\bibfield  {journal}
  {\bibinfo  {journal} {Science Advances}\ }\textbf {\bibinfo {volume} {8}},\
  \bibinfo {pages} {3160} (\bibinfo {year} {2022})},\ \Eprint
  {https://arxiv.org/abs/https://www.science.org/doi/pdf/10.1126/sciadv.abk3160}
  {https://www.science.org/doi/pdf/10.1126/sciadv.abk3160} \BibitemShut
  {NoStop}%
\bibitem [{\citenamefont {Dou}\ and\ \citenamefont
  {Yang}(2023)}]{superconducting_battrey_1}%
  \BibitemOpen
  \bibfield  {author} {\bibinfo {author} {\bibfnamefont {F.-Q.}\ \bibnamefont
  {Dou}}\ and\ \bibinfo {author} {\bibfnamefont {F.-M.}\ \bibnamefont {Yang}},\
  }\bibfield  {title} {\bibinfo {title} {Superconducting transmon
  qubit-resonator quantum battery},\ }\href
  {https://doi.org/10.1103/PhysRevA.107.023725} {\bibfield  {journal} {\bibinfo
   {journal} {Phys. Rev. A}\ }\textbf {\bibinfo {volume} {107}},\ \bibinfo
  {pages} {023725} (\bibinfo {year} {2023})}\BibitemShut {NoStop}%
\bibitem [{\citenamefont {Hu}\ \emph {et~al.}(2022)\citenamefont {Hu},
  \citenamefont {Qiu}, \citenamefont {Souza}, \citenamefont {Yuan},
  \citenamefont {Zhou}, \citenamefont {Zhang}, \citenamefont {Chu},
  \citenamefont {Pan}, \citenamefont {Hu}, \citenamefont {Li}, \citenamefont
  {Xu}, \citenamefont {Zhong}, \citenamefont {Liu}, \citenamefont {Yan},
  \citenamefont {Tan}, \citenamefont {Bachelard}, \citenamefont {Villas-Boas},
  \citenamefont {Santos},\ and\ \citenamefont {Yu}}]{superconductQBexp}%
  \BibitemOpen
  \bibfield  {author} {\bibinfo {author} {\bibfnamefont {C.-K.}\ \bibnamefont
  {Hu}}, \bibinfo {author} {\bibfnamefont {J.}~\bibnamefont {Qiu}}, \bibinfo
  {author} {\bibfnamefont {P.~J.~P.}\ \bibnamefont {Souza}}, \bibinfo {author}
  {\bibfnamefont {J.}~\bibnamefont {Yuan}}, \bibinfo {author} {\bibfnamefont
  {Y.}~\bibnamefont {Zhou}}, \bibinfo {author} {\bibfnamefont {L.}~\bibnamefont
  {Zhang}}, \bibinfo {author} {\bibfnamefont {J.}~\bibnamefont {Chu}}, \bibinfo
  {author} {\bibfnamefont {X.}~\bibnamefont {Pan}}, \bibinfo {author}
  {\bibfnamefont {L.}~\bibnamefont {Hu}}, \bibinfo {author} {\bibfnamefont
  {J.}~\bibnamefont {Li}}, \bibinfo {author} {\bibfnamefont {Y.}~\bibnamefont
  {Xu}}, \bibinfo {author} {\bibfnamefont {Y.}~\bibnamefont {Zhong}}, \bibinfo
  {author} {\bibfnamefont {S.}~\bibnamefont {Liu}}, \bibinfo {author}
  {\bibfnamefont {F.}~\bibnamefont {Yan}}, \bibinfo {author} {\bibfnamefont
  {D.}~\bibnamefont {Tan}}, \bibinfo {author} {\bibfnamefont {R.}~\bibnamefont
  {Bachelard}}, \bibinfo {author} {\bibfnamefont {C.~J.}\ \bibnamefont
  {Villas-Boas}}, \bibinfo {author} {\bibfnamefont {A.~C.}\ \bibnamefont
  {Santos}},\ and\ \bibinfo {author} {\bibfnamefont {D.}~\bibnamefont {Yu}},\
  }\bibfield  {title} {\bibinfo {title} {Optimal charging of a superconducting
  quantum battery},\ }\href {https://doi.org/10.1088/2058-9565/ac8444}
  {\bibfield  {journal} {\bibinfo  {journal} {Quantum Science and Technology}\
  }\textbf {\bibinfo {volume} {7}},\ \bibinfo {pages} {045018} (\bibinfo {year}
  {2022})}\BibitemShut {NoStop}%
\bibitem [{\citenamefont {Gemme}\ \emph {et~al.}(2022)\citenamefont {Gemme},
  \citenamefont {Grossi}, \citenamefont {Ferraro}, \citenamefont {Vallecorsa},\
  and\ \citenamefont {Sassetti}}]{GemmeexpIBMsupercond}%
  \BibitemOpen
  \bibfield  {author} {\bibinfo {author} {\bibfnamefont {G.}~\bibnamefont
  {Gemme}}, \bibinfo {author} {\bibfnamefont {M.}~\bibnamefont {Grossi}},
  \bibinfo {author} {\bibfnamefont {D.}~\bibnamefont {Ferraro}}, \bibinfo
  {author} {\bibfnamefont {S.}~\bibnamefont {Vallecorsa}},\ and\ \bibinfo
  {author} {\bibfnamefont {M.}~\bibnamefont {Sassetti}},\ }\bibfield  {title}
  {\bibinfo {title} {Ibm quantum platforms: A quantum battery perspective},\
  }\bibfield  {journal} {\bibinfo  {journal} {Batteries}\ }\textbf {\bibinfo
  {volume} {8}},\ \href {https://doi.org/10.3390/batteries8050043}
  {10.3390/batteries8050043} (\bibinfo {year} {2022})\BibitemShut {NoStop}%
\bibitem [{\citenamefont {Kurman}\ \emph {et~al.}(2026)\citenamefont {Kurman},
  \citenamefont {Hymas}, \citenamefont {Fedorov}, \citenamefont {Munro},\ and\
  \citenamefont {Quach}}]{kurman_prx_2026}%
  \BibitemOpen
  \bibfield  {author} {\bibinfo {author} {\bibfnamefont {Y.}~\bibnamefont
  {Kurman}}, \bibinfo {author} {\bibfnamefont {K.}~\bibnamefont {Hymas}},
  \bibinfo {author} {\bibfnamefont {A.}~\bibnamefont {Fedorov}}, \bibinfo
  {author} {\bibfnamefont {W.~J.}\ \bibnamefont {Munro}},\ and\ \bibinfo
  {author} {\bibfnamefont {J.}~\bibnamefont {Quach}},\ }\bibfield  {title}
  {\bibinfo {title} {Powering quantum computation with quantum batteries},\
  }\href {https://doi.org/10.1103/l39v-jwwz} {\bibfield  {journal} {\bibinfo
  {journal} {Phys. Rev. X}\ }\textbf {\bibinfo {volume} {16}},\ \bibinfo
  {pages} {011016} (\bibinfo {year} {2026})}\BibitemShut {NoStop}%
\bibitem [{\citenamefont {Gyhm}\ \emph {et~al.}(2022)\citenamefont {Gyhm},
  \citenamefont {\ifmmode~\check{S}\else \v{S}\fi{}afr\'anek},\ and\
  \citenamefont {Rosa}}]{gyhm_prl_2022}%
  \BibitemOpen
  \bibfield  {author} {\bibinfo {author} {\bibfnamefont {J.-Y.}\ \bibnamefont
  {Gyhm}}, \bibinfo {author} {\bibfnamefont {D.}~\bibnamefont
  {\ifmmode~\check{S}\else \v{S}\fi{}afr\'anek}},\ and\ \bibinfo {author}
  {\bibfnamefont {D.}~\bibnamefont {Rosa}},\ }\bibfield  {title} {\bibinfo
  {title} {Quantum charging advantage cannot be extensive without global
  operations},\ }\href {https://doi.org/10.1103/PhysRevLett.128.140501}
  {\bibfield  {journal} {\bibinfo  {journal} {Phys. Rev. Lett.}\ }\textbf
  {\bibinfo {volume} {128}},\ \bibinfo {pages} {140501} (\bibinfo {year}
  {2022})}\BibitemShut {NoStop}%
\bibitem [{\citenamefont {Horodecki}\ \emph {et~al.}(2009)\citenamefont
  {Horodecki}, \citenamefont {Horodecki}, \citenamefont {Horodecki},\ and\
  \citenamefont {Horodecki}}]{horodecki_review}%
  \BibitemOpen
  \bibfield  {author} {\bibinfo {author} {\bibfnamefont {R.}~\bibnamefont
  {Horodecki}}, \bibinfo {author} {\bibfnamefont {P.}~\bibnamefont
  {Horodecki}}, \bibinfo {author} {\bibfnamefont {M.}~\bibnamefont
  {Horodecki}},\ and\ \bibinfo {author} {\bibfnamefont {K.}~\bibnamefont
  {Horodecki}},\ }\bibfield  {title} {\bibinfo {title} {Quantum entanglement},\
  }\href {https://doi.org/10.1103/RevModPhys.81.865} {\bibfield  {journal}
  {\bibinfo  {journal} {Rev. Mod. Phys.}\ }\textbf {\bibinfo {volume} {81}},\
  \bibinfo {pages} {865} (\bibinfo {year} {2009})}\BibitemShut {NoStop}%
\bibitem [{\citenamefont {Aaronson}\ and\ \citenamefont
  {Gottesman}(2004)}]{aaronson_pra_2004}%
  \BibitemOpen
  \bibfield  {author} {\bibinfo {author} {\bibfnamefont {S.}~\bibnamefont
  {Aaronson}}\ and\ \bibinfo {author} {\bibfnamefont {D.}~\bibnamefont
  {Gottesman}},\ }\bibfield  {title} {\bibinfo {title} {Improved simulation of
  stabilizer circuits},\ }\href {https://doi.org/10.1103/PhysRevA.70.052328}
  {\bibfield  {journal} {\bibinfo  {journal} {Phys. Rev. A}\ }\textbf {\bibinfo
  {volume} {70}},\ \bibinfo {pages} {052328} (\bibinfo {year}
  {2004})}\BibitemShut {NoStop}%
\bibitem [{\citenamefont {Nielsen}\ and\ \citenamefont
  {Chuang}(2010)}]{nielsen2010}%
  \BibitemOpen
  \bibfield  {author} {\bibinfo {author} {\bibfnamefont {M.~A.}\ \bibnamefont
  {Nielsen}}\ and\ \bibinfo {author} {\bibfnamefont {I.~L.}\ \bibnamefont
  {Chuang}},\ }\href@noop {} {\emph {\bibinfo {title} {Quantum computation and
  quantum information}}}\ (\bibinfo  {publisher} {Cambridge university press},\
  \bibinfo {year} {2010})\BibitemShut {NoStop}%
\bibitem [{\citenamefont {Bravyi}\ and\ \citenamefont
  {Kitaev}(2005)}]{bravyi_pra_2005}%
  \BibitemOpen
  \bibfield  {author} {\bibinfo {author} {\bibfnamefont {S.}~\bibnamefont
  {Bravyi}}\ and\ \bibinfo {author} {\bibfnamefont {A.}~\bibnamefont
  {Kitaev}},\ }\bibfield  {title} {\bibinfo {title} {Universal quantum
  computation with ideal clifford gates and noisy ancillas},\ }\href
  {https://doi.org/10.1103/PhysRevA.71.022316} {\bibfield  {journal} {\bibinfo
  {journal} {Phys. Rev. A}\ }\textbf {\bibinfo {volume} {71}},\ \bibinfo
  {pages} {022316} (\bibinfo {year} {2005})}\BibitemShut {NoStop}%
\bibitem [{\citenamefont {Gottesman}(1998{\natexlab{a}})}]{gottesman1998}%
  \BibitemOpen
  \bibfield  {author} {\bibinfo {author} {\bibfnamefont {D.}~\bibnamefont
  {Gottesman}},\ }\href {https://arxiv.org/abs/quant-ph/9807006} {\bibinfo
  {title} {The heisenberg representation of quantum computers}} (\bibinfo
  {year} {1998}{\natexlab{a}}),\ \Eprint
  {https://arxiv.org/abs/quant-ph/9807006} {arXiv:quant-ph/9807006 [quant-ph]}
  \BibitemShut {NoStop}%
\bibitem [{\citenamefont {Gottesman}(1997)}]{gottesman1997}%
  \BibitemOpen
  \bibfield  {author} {\bibinfo {author} {\bibfnamefont {D.}~\bibnamefont
  {Gottesman}},\ }\href {https://arxiv.org/abs/quant-ph/9705052} {\bibinfo
  {title} {Stabilizer codes and quantum error correction}} (\bibinfo {year}
  {1997}),\ \Eprint {https://arxiv.org/abs/quant-ph/9705052}
  {arXiv:quant-ph/9705052 [quant-ph]} \BibitemShut {NoStop}%
\bibitem [{\citenamefont {Gottesman}(1998{\natexlab{b}})}]{gottesman_pra_1998}%
  \BibitemOpen
  \bibfield  {author} {\bibinfo {author} {\bibfnamefont {D.}~\bibnamefont
  {Gottesman}},\ }\bibfield  {title} {\bibinfo {title} {Theory of
  fault-tolerant quantum computation},\ }\href
  {https://doi.org/10.1103/PhysRevA.57.127} {\bibfield  {journal} {\bibinfo
  {journal} {Phys. Rev. A}\ }\textbf {\bibinfo {volume} {57}},\ \bibinfo
  {pages} {127} (\bibinfo {year} {1998}{\natexlab{b}})}\BibitemShut {NoStop}%
\bibitem [{\citenamefont {Howard}\ and\ \citenamefont
  {Campbell}(2017)}]{howard_prl_2017}%
  \BibitemOpen
  \bibfield  {author} {\bibinfo {author} {\bibfnamefont {M.}~\bibnamefont
  {Howard}}\ and\ \bibinfo {author} {\bibfnamefont {E.}~\bibnamefont
  {Campbell}},\ }\bibfield  {title} {\bibinfo {title} {Application of a
  resource theory for magic states to fault-tolerant quantum computing},\
  }\href {https://doi.org/10.1103/PhysRevLett.118.090501} {\bibfield  {journal}
  {\bibinfo  {journal} {Phys. Rev. Lett.}\ }\textbf {\bibinfo {volume} {118}},\
  \bibinfo {pages} {090501} (\bibinfo {year} {2017})}\BibitemShut {NoStop}%
\bibitem [{\citenamefont {Leone}\ \emph {et~al.}(2022)\citenamefont {Leone},
  \citenamefont {Oliviero},\ and\ \citenamefont {Hamma}}]{leone_prl_2022}%
  \BibitemOpen
  \bibfield  {author} {\bibinfo {author} {\bibfnamefont {L.}~\bibnamefont
  {Leone}}, \bibinfo {author} {\bibfnamefont {S.~F.~E.}\ \bibnamefont
  {Oliviero}},\ and\ \bibinfo {author} {\bibfnamefont {A.}~\bibnamefont
  {Hamma}},\ }\bibfield  {title} {\bibinfo {title} {Stabilizer r\'enyi
  entropy},\ }\href {https://doi.org/10.1103/PhysRevLett.128.050402} {\bibfield
   {journal} {\bibinfo  {journal} {Phys. Rev. Lett.}\ }\textbf {\bibinfo
  {volume} {128}},\ \bibinfo {pages} {050402} (\bibinfo {year}
  {2022})}\BibitemShut {NoStop}%
\bibitem [{\citenamefont {White}\ \emph {et~al.}(2021)\citenamefont {White},
  \citenamefont {Cao},\ and\ \citenamefont {Swingle}}]{white_prb_2021}%
  \BibitemOpen
  \bibfield  {author} {\bibinfo {author} {\bibfnamefont {C.~D.}\ \bibnamefont
  {White}}, \bibinfo {author} {\bibfnamefont {C.}~\bibnamefont {Cao}},\ and\
  \bibinfo {author} {\bibfnamefont {B.}~\bibnamefont {Swingle}},\ }\bibfield
  {title} {\bibinfo {title} {Conformal field theories are magical},\ }\href
  {https://doi.org/10.1103/PhysRevB.103.075145} {\bibfield  {journal} {\bibinfo
   {journal} {Phys. Rev. B}\ }\textbf {\bibinfo {volume} {103}},\ \bibinfo
  {pages} {075145} (\bibinfo {year} {2021})}\BibitemShut {NoStop}%
\bibitem [{\citenamefont {Sarkar}\ \emph {et~al.}(2020)\citenamefont {Sarkar},
  \citenamefont {Mukhopadhyay},\ and\ \citenamefont {Bayat}}]{Sarkar2020}%
  \BibitemOpen
  \bibfield  {author} {\bibinfo {author} {\bibfnamefont {S.}~\bibnamefont
  {Sarkar}}, \bibinfo {author} {\bibfnamefont {C.}~\bibnamefont
  {Mukhopadhyay}},\ and\ \bibinfo {author} {\bibfnamefont {A.}~\bibnamefont
  {Bayat}},\ }\bibfield  {title} {\bibinfo {title} {Characterization of an
  operational quantum resource in a critical many-body system},\ }\href
  {https://doi.org/10.1088/1367-2630/aba919} {\bibfield  {journal} {\bibinfo
  {journal} {New Journal of Physics}\ }\textbf {\bibinfo {volume} {22}},\
  \bibinfo {pages} {083077} (\bibinfo {year} {2020})}\BibitemShut {NoStop}%
\bibitem [{\citenamefont {Oliviero}\ \emph
  {et~al.}(2022{\natexlab{a}})\citenamefont {Oliviero}, \citenamefont {Leone},\
  and\ \citenamefont {Hamma}}]{oliviero_pra_2022}%
  \BibitemOpen
  \bibfield  {author} {\bibinfo {author} {\bibfnamefont {S.~F.~E.}\
  \bibnamefont {Oliviero}}, \bibinfo {author} {\bibfnamefont {L.}~\bibnamefont
  {Leone}},\ and\ \bibinfo {author} {\bibfnamefont {A.}~\bibnamefont {Hamma}},\
  }\bibfield  {title} {\bibinfo {title} {Magic-state resource theory for the
  ground state of the transverse-field ising model},\ }\href
  {https://doi.org/10.1103/PhysRevA.106.042426} {\bibfield  {journal} {\bibinfo
   {journal} {Phys. Rev. A}\ }\textbf {\bibinfo {volume} {106}},\ \bibinfo
  {pages} {042426} (\bibinfo {year} {2022}{\natexlab{a}})}\BibitemShut
  {NoStop}%
\bibitem [{\citenamefont {Haug}\ and\ \citenamefont
  {Piroli}(2023)}]{haug_prb_2023}%
  \BibitemOpen
  \bibfield  {author} {\bibinfo {author} {\bibfnamefont {T.}~\bibnamefont
  {Haug}}\ and\ \bibinfo {author} {\bibfnamefont {L.}~\bibnamefont {Piroli}},\
  }\bibfield  {title} {\bibinfo {title} {Quantifying nonstabilizerness of
  matrix product states},\ }\href {https://doi.org/10.1103/PhysRevB.107.035148}
  {\bibfield  {journal} {\bibinfo  {journal} {Phys. Rev. B}\ }\textbf {\bibinfo
  {volume} {107}},\ \bibinfo {pages} {035148} (\bibinfo {year}
  {2023})}\BibitemShut {NoStop}%
\bibitem [{\citenamefont {Tarabunga}\ \emph {et~al.}(2023)\citenamefont
  {Tarabunga}, \citenamefont {Tirrito}, \citenamefont {Chanda},\ and\
  \citenamefont {Dalmonte}}]{tarabunga_prxq_2023}%
  \BibitemOpen
  \bibfield  {author} {\bibinfo {author} {\bibfnamefont {P.~S.}\ \bibnamefont
  {Tarabunga}}, \bibinfo {author} {\bibfnamefont {E.}~\bibnamefont {Tirrito}},
  \bibinfo {author} {\bibfnamefont {T.}~\bibnamefont {Chanda}},\ and\ \bibinfo
  {author} {\bibfnamefont {M.}~\bibnamefont {Dalmonte}},\ }\bibfield  {title}
  {\bibinfo {title} {Many-body magic via pauli-markov chains---from criticality
  to gauge theories},\ }\href {https://doi.org/10.1103/PRXQuantum.4.040317}
  {\bibfield  {journal} {\bibinfo  {journal} {PRX Quantum}\ }\textbf {\bibinfo
  {volume} {4}},\ \bibinfo {pages} {040317} (\bibinfo {year}
  {2023})}\BibitemShut {NoStop}%
\bibitem [{\citenamefont {Tarabunga}(2024)}]{Tarabunga2024}%
  \BibitemOpen
  \bibfield  {author} {\bibinfo {author} {\bibfnamefont {P.~S.}\ \bibnamefont
  {Tarabunga}},\ }\bibfield  {title} {\bibinfo {title} {Critical behaviors of
  non-stabilizerness in quantum spin chains},\ }\href
  {https://doi.org/10.22331/q-2024-07-17-1413} {\bibfield  {journal} {\bibinfo
  {journal} {{Quantum}}\ }\textbf {\bibinfo {volume} {8}},\ \bibinfo {pages}
  {1413} (\bibinfo {year} {2024})}\BibitemShut {NoStop}%
\bibitem [{\citenamefont {Grabarits}\ and\ \citenamefont {del
  Campo}(2026)}]{grabarits2026}%
  \BibitemOpen
  \bibfield  {author} {\bibinfo {author} {\bibfnamefont {A.}~\bibnamefont
  {Grabarits}}\ and\ \bibinfo {author} {\bibfnamefont {A.}~\bibnamefont {del
  Campo}},\ }\href {https://arxiv.org/abs/2603.08841} {\bibinfo {title}
  {Universal non-stabilizerness dynamics across quantum phase transitions}}
  (\bibinfo {year} {2026}),\ \Eprint {https://arxiv.org/abs/2603.08841}
  {arXiv:2603.08841 [quant-ph]} \BibitemShut {NoStop}%
\bibitem [{\citenamefont {Falc\~ao}\ \emph {et~al.}(2025)\citenamefont
  {Falc\~ao}, \citenamefont {Sierant}, \citenamefont {Zakrzewski},\ and\
  \citenamefont {Tirrito}}]{falcao_prl_2025}%
  \BibitemOpen
  \bibfield  {author} {\bibinfo {author} {\bibfnamefont {P.~R.~N.}\
  \bibnamefont {Falc\~ao}}, \bibinfo {author} {\bibfnamefont {P.}~\bibnamefont
  {Sierant}}, \bibinfo {author} {\bibfnamefont {J.}~\bibnamefont
  {Zakrzewski}},\ and\ \bibinfo {author} {\bibfnamefont {E.}~\bibnamefont
  {Tirrito}},\ }\bibfield  {title} {\bibinfo {title} {Nonstabilizerness
  dynamics in many-body localized systems},\ }\href
  {https://doi.org/10.1103/xfp5-hhs4} {\bibfield  {journal} {\bibinfo
  {journal} {Phys. Rev. Lett.}\ }\textbf {\bibinfo {volume} {135}},\ \bibinfo
  {pages} {240404} (\bibinfo {year} {2025})}\BibitemShut {NoStop}%
\bibitem [{\citenamefont {Li}\ \emph {et~al.}(2026{\natexlab{a}})\citenamefont
  {Li}, \citenamefont {Zhang}, \citenamefont {Zhao}, \citenamefont {Huang},\
  and\ \citenamefont {Zhong}}]{li2026}%
  \BibitemOpen
  \bibfield  {author} {\bibinfo {author} {\bibfnamefont {H.-Z.}\ \bibnamefont
  {Li}}, \bibinfo {author} {\bibfnamefont {Y.-R.}\ \bibnamefont {Zhang}},
  \bibinfo {author} {\bibfnamefont {Y.-J.}\ \bibnamefont {Zhao}}, \bibinfo
  {author} {\bibfnamefont {X.}~\bibnamefont {Huang}},\ and\ \bibinfo {author}
  {\bibfnamefont {J.-X.}\ \bibnamefont {Zhong}},\ }\bibfield  {title} {\bibinfo
  {title} {Slow growth of quantum nonstabilizerness in disorder-free stark
  many-body localization},\ }\href {https://doi.org/10.1103/195d-r5j3}
  {\bibfield  {journal} {\bibinfo  {journal} {Phys. Rev. B}\ }\textbf {\bibinfo
  {volume} {113}},\ \bibinfo {pages} {104305} (\bibinfo {year}
  {2026}{\natexlab{a}})}\BibitemShut {NoStop}%
\bibitem [{\citenamefont {Campbell}\ \emph {et~al.}(2017)\citenamefont
  {Campbell}, \citenamefont {Terhal},\ and\ \citenamefont
  {Vuillot}}]{Campbell2017}%
  \BibitemOpen
  \bibfield  {author} {\bibinfo {author} {\bibfnamefont {E.~T.}\ \bibnamefont
  {Campbell}}, \bibinfo {author} {\bibfnamefont {B.~M.}\ \bibnamefont
  {Terhal}},\ and\ \bibinfo {author} {\bibfnamefont {C.}~\bibnamefont
  {Vuillot}},\ }\bibfield  {title} {\bibinfo {title} {Roads towards
  fault-tolerant universal quantum computation},\ }\href
  {https://doi.org/10.1038/nature23460} {\bibfield  {journal} {\bibinfo
  {journal} {Nature}\ }\textbf {\bibinfo {volume} {549}},\ \bibinfo {pages}
  {172–179} (\bibinfo {year} {2017})}\BibitemShut {NoStop}%
\bibitem [{\citenamefont {Bravyi}\ \emph {et~al.}(2019)\citenamefont {Bravyi},
  \citenamefont {Browne}, \citenamefont {Calpin}, \citenamefont {Campbell},
  \citenamefont {Gosset},\ and\ \citenamefont {Howard}}]{Bravyi2019}%
  \BibitemOpen
  \bibfield  {author} {\bibinfo {author} {\bibfnamefont {S.}~\bibnamefont
  {Bravyi}}, \bibinfo {author} {\bibfnamefont {D.}~\bibnamefont {Browne}},
  \bibinfo {author} {\bibfnamefont {P.}~\bibnamefont {Calpin}}, \bibinfo
  {author} {\bibfnamefont {E.}~\bibnamefont {Campbell}}, \bibinfo {author}
  {\bibfnamefont {D.}~\bibnamefont {Gosset}},\ and\ \bibinfo {author}
  {\bibfnamefont {M.}~\bibnamefont {Howard}},\ }\bibfield  {title} {\bibinfo
  {title} {Simulation of quantum circuits by low-rank stabilizer
  decompositions},\ }\href {https://doi.org/10.22331/q-2019-09-02-181}
  {\bibfield  {journal} {\bibinfo  {journal} {{Quantum}}\ }\textbf {\bibinfo
  {volume} {3}},\ \bibinfo {pages} {181} (\bibinfo {year} {2019})}\BibitemShut
  {NoStop}%
\bibitem [{\citenamefont {Oliviero}\ \emph
  {et~al.}(2022{\natexlab{b}})\citenamefont {Oliviero}, \citenamefont {Leone},
  \citenamefont {Hamma},\ and\ \citenamefont {Lloyd}}]{Oliviero2022}%
  \BibitemOpen
  \bibfield  {author} {\bibinfo {author} {\bibfnamefont {S.~F.~E.}\
  \bibnamefont {Oliviero}}, \bibinfo {author} {\bibfnamefont {L.}~\bibnamefont
  {Leone}}, \bibinfo {author} {\bibfnamefont {A.}~\bibnamefont {Hamma}},\ and\
  \bibinfo {author} {\bibfnamefont {S.}~\bibnamefont {Lloyd}},\ }\bibfield
  {title} {\bibinfo {title} {Measuring magic on a quantum processor},\
  }\bibfield  {journal} {\bibinfo  {journal} {npj Quantum Information}\
  }\textbf {\bibinfo {volume} {8}},\ \href
  {https://doi.org/10.1038/s41534-022-00666-5} {10.1038/s41534-022-00666-5}
  (\bibinfo {year} {2022}{\natexlab{b}})\BibitemShut {NoStop}%
\bibitem [{\citenamefont {Capecci}\ \emph {et~al.}(2025)\citenamefont
  {Capecci}, \citenamefont {Santra}, \citenamefont {Bottarelli}, \citenamefont
  {Tirrito},\ and\ \citenamefont {Hauke}}]{capecci2025}%
  \BibitemOpen
  \bibfield  {author} {\bibinfo {author} {\bibfnamefont {C.}~\bibnamefont
  {Capecci}}, \bibinfo {author} {\bibfnamefont {G.~C.}\ \bibnamefont {Santra}},
  \bibinfo {author} {\bibfnamefont {A.}~\bibnamefont {Bottarelli}}, \bibinfo
  {author} {\bibfnamefont {E.}~\bibnamefont {Tirrito}},\ and\ \bibinfo {author}
  {\bibfnamefont {P.}~\bibnamefont {Hauke}},\ }\href
  {https://arxiv.org/abs/2505.17185} {\bibinfo {title} {Role of
  nonstabilizerness in quantum optimization}} (\bibinfo {year} {2025}),\
  \Eprint {https://arxiv.org/abs/2505.17185} {arXiv:2505.17185 [quant-ph]}
  \BibitemShut {NoStop}%
\bibitem [{\citenamefont {Hern\'andez-Yanes}\ \emph {et~al.}(2026)\citenamefont
  {Hern\'andez-Yanes}, \citenamefont {Sierant}, \citenamefont {Zakrzewski},\
  and\ \citenamefont {P\l{}odzie\ifmmode~\acute{n}\else
  \'{n}\fi{}}}]{yanes_pra_2026}%
  \BibitemOpen
  \bibfield  {author} {\bibinfo {author} {\bibfnamefont {T.}~\bibnamefont
  {Hern\'andez-Yanes}}, \bibinfo {author} {\bibfnamefont {P.}~\bibnamefont
  {Sierant}}, \bibinfo {author} {\bibfnamefont {J.}~\bibnamefont
  {Zakrzewski}},\ and\ \bibinfo {author} {\bibfnamefont {M.}~\bibnamefont
  {P\l{}odzie\ifmmode~\acute{n}\else \'{n}\fi{}}},\ }\bibfield  {title}
  {\bibinfo {title} {Nonstabilizerness in quantum-enhanced metrological
  protocols},\ }\href {https://doi.org/10.1103/tmf9-fyc2} {\bibfield  {journal}
  {\bibinfo  {journal} {Phys. Rev. A}\ }\textbf {\bibinfo {volume} {113}},\
  \bibinfo {pages} {012416} (\bibinfo {year} {2026})}\BibitemShut {NoStop}%
\bibitem [{\citenamefont {Veitch}\ \emph {et~al.}(2014)\citenamefont {Veitch},
  \citenamefont {Hamed~Mousavian}, \citenamefont {Gottesman},\ and\
  \citenamefont {Emerson}}]{Veitch2014}%
  \BibitemOpen
  \bibfield  {author} {\bibinfo {author} {\bibfnamefont {V.}~\bibnamefont
  {Veitch}}, \bibinfo {author} {\bibfnamefont {S.~A.}\ \bibnamefont
  {Hamed~Mousavian}}, \bibinfo {author} {\bibfnamefont {D.}~\bibnamefont
  {Gottesman}},\ and\ \bibinfo {author} {\bibfnamefont {J.}~\bibnamefont
  {Emerson}},\ }\bibfield  {title} {\bibinfo {title} {The resource theory of
  stabilizer quantum computation},\ }\href
  {https://doi.org/10.1088/1367-2630/16/1/013009} {\bibfield  {journal}
  {\bibinfo  {journal} {New Journal of Physics}\ }\textbf {\bibinfo {volume}
  {16}},\ \bibinfo {pages} {013009} (\bibinfo {year} {2014})}\BibitemShut
  {NoStop}%
\bibitem [{\citenamefont {Sierant}\ \emph {et~al.}(2026)\citenamefont
  {Sierant}, \citenamefont {Vall{\`{e}}s-Muns},\ and\ \citenamefont
  {Garcia-Saez}}]{Sierant2026}%
  \BibitemOpen
  \bibfield  {author} {\bibinfo {author} {\bibfnamefont {P.}~\bibnamefont
  {Sierant}}, \bibinfo {author} {\bibfnamefont {J.}~\bibnamefont
  {Vall{\`{e}}s-Muns}},\ and\ \bibinfo {author} {\bibfnamefont
  {A.}~\bibnamefont {Garcia-Saez}},\ }\bibfield  {title} {\bibinfo {title}
  {Computing quantum magic of state vectors},\ }\href
  {https://doi.org/10.22331/q-2026-04-10-2059} {\bibfield  {journal} {\bibinfo
  {journal} {{Quantum}}\ }\textbf {\bibinfo {volume} {10}},\ \bibinfo {pages}
  {2059} (\bibinfo {year} {2026})}\BibitemShut {NoStop}%
\bibitem [{\citenamefont {Allahverdyan}\ \emph {et~al.}(2004)\citenamefont
  {Allahverdyan}, \citenamefont {Balian},\ and\ \citenamefont
  {Nieuwenhuizen}}]{Allahverdyan2004}%
  \BibitemOpen
  \bibfield  {author} {\bibinfo {author} {\bibfnamefont {A.~E.}\ \bibnamefont
  {Allahverdyan}}, \bibinfo {author} {\bibfnamefont {R.}~\bibnamefont
  {Balian}},\ and\ \bibinfo {author} {\bibfnamefont {T.~M.}\ \bibnamefont
  {Nieuwenhuizen}},\ }\bibfield  {title} {\bibinfo {title} {Maximal work
  extraction from finite quantum systems},\ }\href
  {https://doi.org/10.1209/epl/i2004-10101-2} {\bibfield  {journal} {\bibinfo
  {journal} {Europhysics Letters (EPL)}\ }\textbf {\bibinfo {volume} {67}},\
  \bibinfo {pages} {565–571} (\bibinfo {year} {2004})}\BibitemShut {NoStop}%
\bibitem [{\citenamefont {Tirrito}\ \emph
  {et~al.}(2025{\natexlab{a}})\citenamefont {Tirrito}, \citenamefont
  {Tarabunga}, \citenamefont {Bhakuni}, \citenamefont {Dalmonte}, \citenamefont
  {Sierant},\ and\ \citenamefont {Turkeshi}}]{tirrito2025}%
  \BibitemOpen
  \bibfield  {author} {\bibinfo {author} {\bibfnamefont {E.}~\bibnamefont
  {Tirrito}}, \bibinfo {author} {\bibfnamefont {P.~S.}\ \bibnamefont
  {Tarabunga}}, \bibinfo {author} {\bibfnamefont {D.~S.}\ \bibnamefont
  {Bhakuni}}, \bibinfo {author} {\bibfnamefont {M.}~\bibnamefont {Dalmonte}},
  \bibinfo {author} {\bibfnamefont {P.}~\bibnamefont {Sierant}},\ and\ \bibinfo
  {author} {\bibfnamefont {X.}~\bibnamefont {Turkeshi}},\ }\href
  {https://arxiv.org/abs/2506.12133} {\bibinfo {title} {Universal spreading of
  nonstabilizerness and quantum transport}} (\bibinfo {year}
  {2025}{\natexlab{a}}),\ \Eprint {https://arxiv.org/abs/2506.12133}
  {arXiv:2506.12133 [quant-ph]} \BibitemShut {NoStop}%
\bibitem [{\citenamefont {Fehske}\ and\ \citenamefont
  {Schneider}(2008)}]{Fehske08}%
  \BibitemOpen
  \bibfield  {author} {\bibinfo {author} {\bibfnamefont {H.}~\bibnamefont
  {Fehske}}\ and\ \bibinfo {author} {\bibfnamefont {R.}~\bibnamefont
  {Schneider}},\ }\href {http://dx.doi.org/10.1007/978-3-540-74686-7} {\emph
  {\bibinfo {title} {Computational many-particle physics}}}\ (\bibinfo
  {publisher} {Springer},\ \bibinfo {address} {Germany},\ \bibinfo {year}
  {2008})\BibitemShut {NoStop}%
\bibitem [{\citenamefont {Sachdev}\ and\ \citenamefont
  {Ye}(1993)}]{sachdev_prl_1993}%
  \BibitemOpen
  \bibfield  {author} {\bibinfo {author} {\bibfnamefont {S.}~\bibnamefont
  {Sachdev}}\ and\ \bibinfo {author} {\bibfnamefont {J.}~\bibnamefont {Ye}},\
  }\bibfield  {title} {\bibinfo {title} {Gapless spin-fluid ground state in a
  random quantum heisenberg magnet},\ }\href
  {https://doi.org/10.1103/PhysRevLett.70.3339} {\bibfield  {journal} {\bibinfo
   {journal} {Phys. Rev. Lett.}\ }\textbf {\bibinfo {volume} {70}},\ \bibinfo
  {pages} {3339} (\bibinfo {year} {1993})}\BibitemShut {NoStop}%
\bibitem [{\citenamefont {Bera}\ and\ \citenamefont
  {Schiro}(2025)}]{bera_scipost_2025}%
  \BibitemOpen
  \bibfield  {author} {\bibinfo {author} {\bibfnamefont {S.}~\bibnamefont
  {Bera}}\ and\ \bibinfo {author} {\bibfnamefont {M.}~\bibnamefont {Schiro}},\
  }\bibfield  {title} {\bibinfo {title} {{Non-stabilizerness of
  Sachdev-Ye-Kitaev model}},\ }\href
  {https://doi.org/10.21468/SciPostPhys.19.6.159} {\bibfield  {journal}
  {\bibinfo  {journal} {SciPost Phys.}\ }\textbf {\bibinfo {volume} {19}},\
  \bibinfo {pages} {159} (\bibinfo {year} {2025})}\BibitemShut {NoStop}%
\bibitem [{\citenamefont {Russomanno}\ \emph {et~al.}(2025)\citenamefont
  {Russomanno}, \citenamefont {Passarelli}, \citenamefont {Rossini},\ and\
  \citenamefont {Lucignano}}]{russomanno_prb_2025}%
  \BibitemOpen
  \bibfield  {author} {\bibinfo {author} {\bibfnamefont {A.}~\bibnamefont
  {Russomanno}}, \bibinfo {author} {\bibfnamefont {G.}~\bibnamefont
  {Passarelli}}, \bibinfo {author} {\bibfnamefont {D.}~\bibnamefont
  {Rossini}},\ and\ \bibinfo {author} {\bibfnamefont {P.}~\bibnamefont
  {Lucignano}},\ }\bibfield  {title} {\bibinfo {title} {Nonstabilizerness in
  the unitary and monitored quantum dynamics of xxz-staggered and
  sachdev-ye-kitaev models},\ }\href {https://doi.org/10.1103/njgn-fksh}
  {\bibfield  {journal} {\bibinfo  {journal} {Phys. Rev. B}\ }\textbf {\bibinfo
  {volume} {112}},\ \bibinfo {pages} {064312} (\bibinfo {year}
  {2025})}\BibitemShut {NoStop}%
\bibitem [{\citenamefont {Jasser}\ \emph {et~al.}(2025)\citenamefont {Jasser},
  \citenamefont {Odavi\ifmmode~\acute{c}\else \'{c}\fi{}},\ and\ \citenamefont
  {Hamma}}]{jasser_prb_2025}%
  \BibitemOpen
  \bibfield  {author} {\bibinfo {author} {\bibfnamefont {B.}~\bibnamefont
  {Jasser}}, \bibinfo {author} {\bibfnamefont {J.}~\bibnamefont
  {Odavi\ifmmode~\acute{c}\else \'{c}\fi{}}},\ and\ \bibinfo {author}
  {\bibfnamefont {A.}~\bibnamefont {Hamma}},\ }\bibfield  {title} {\bibinfo
  {title} {Stabilizer entropy and entanglement complexity in the
  sachdev-ye-kitaev model},\ }\href {https://doi.org/10.1103/rz86-47h3}
  {\bibfield  {journal} {\bibinfo  {journal} {Phys. Rev. B}\ }\textbf {\bibinfo
  {volume} {112}},\ \bibinfo {pages} {174204} (\bibinfo {year}
  {2025})}\BibitemShut {NoStop}%
\bibitem [{\citenamefont {Rossini}\ \emph {et~al.}(2020)\citenamefont
  {Rossini}, \citenamefont {Andolina}, \citenamefont {Rosa}, \citenamefont
  {Carrega},\ and\ \citenamefont {Polini}}]{rossini_prl_2020}%
  \BibitemOpen
  \bibfield  {author} {\bibinfo {author} {\bibfnamefont {D.}~\bibnamefont
  {Rossini}}, \bibinfo {author} {\bibfnamefont {G.~M.}\ \bibnamefont
  {Andolina}}, \bibinfo {author} {\bibfnamefont {D.}~\bibnamefont {Rosa}},
  \bibinfo {author} {\bibfnamefont {M.}~\bibnamefont {Carrega}},\ and\ \bibinfo
  {author} {\bibfnamefont {M.}~\bibnamefont {Polini}},\ }\bibfield  {title}
  {\bibinfo {title} {Quantum advantage in the charging process of
  sachdev-ye-kitaev batteries},\ }\href
  {https://doi.org/10.1103/PhysRevLett.125.236402} {\bibfield  {journal}
  {\bibinfo  {journal} {Phys. Rev. Lett.}\ }\textbf {\bibinfo {volume} {125}},\
  \bibinfo {pages} {236402} (\bibinfo {year} {2020})}\BibitemShut {NoStop}%
\bibitem [{\citenamefont {Rosa}\ \emph {et~al.}(2020)\citenamefont {Rosa},
  \citenamefont {Rossini}, \citenamefont {Andolina}, \citenamefont {Polini},\
  and\ \citenamefont {Carrega}}]{Rosa2020}%
  \BibitemOpen
  \bibfield  {author} {\bibinfo {author} {\bibfnamefont {D.}~\bibnamefont
  {Rosa}}, \bibinfo {author} {\bibfnamefont {D.}~\bibnamefont {Rossini}},
  \bibinfo {author} {\bibfnamefont {G.~M.}\ \bibnamefont {Andolina}}, \bibinfo
  {author} {\bibfnamefont {M.}~\bibnamefont {Polini}},\ and\ \bibinfo {author}
  {\bibfnamefont {M.}~\bibnamefont {Carrega}},\ }\bibfield  {title} {\bibinfo
  {title} {Ultra-stable charging of fast-scrambling syk quantum batteries},\
  }\bibfield  {journal} {\bibinfo  {journal} {Journal of High Energy Physics}\
  }\textbf {\bibinfo {volume} {2020}},\ \href
  {https://doi.org/10.1007/jhep11(2020)067} {10.1007/jhep11(2020)067} (\bibinfo
  {year} {2020})\BibitemShut {NoStop}%
\bibitem [{\citenamefont {Iannotti}\ \emph {et~al.}(2026)\citenamefont
  {Iannotti}, \citenamefont {Russotto}, \citenamefont {Jasser}, \citenamefont
  {Odavić},\ and\ \citenamefont {Hamma}}]{iannotti2026}%
  \BibitemOpen
  \bibfield  {author} {\bibinfo {author} {\bibfnamefont {D.}~\bibnamefont
  {Iannotti}}, \bibinfo {author} {\bibfnamefont {A.}~\bibnamefont {Russotto}},
  \bibinfo {author} {\bibfnamefont {B.}~\bibnamefont {Jasser}}, \bibinfo
  {author} {\bibfnamefont {J.}~\bibnamefont {Odavić}},\ and\ \bibinfo {author}
  {\bibfnamefont {A.}~\bibnamefont {Hamma}},\ }\href
  {https://arxiv.org/abs/2603.28870} {\bibinfo {title} {Non-stabilizerness and
  u(1) symmetry in chaotic many-body quantum systems}} (\bibinfo {year}
  {2026}),\ \Eprint {https://arxiv.org/abs/2603.28870} {arXiv:2603.28870
  [quant-ph]} \BibitemShut {NoStop}%
\bibitem [{\citenamefont {Turkeshi}\ \emph {et~al.}(2025)\citenamefont
  {Turkeshi}, \citenamefont {Dymarsky},\ and\ \citenamefont
  {Sierant}}]{turkeshi_prb_2025}%
  \BibitemOpen
  \bibfield  {author} {\bibinfo {author} {\bibfnamefont {X.}~\bibnamefont
  {Turkeshi}}, \bibinfo {author} {\bibfnamefont {A.}~\bibnamefont {Dymarsky}},\
  and\ \bibinfo {author} {\bibfnamefont {P.}~\bibnamefont {Sierant}},\
  }\bibfield  {title} {\bibinfo {title} {Pauli spectrum and nonstabilizerness
  of typical quantum many-body states},\ }\href
  {https://doi.org/10.1103/PhysRevB.111.054301} {\bibfield  {journal} {\bibinfo
   {journal} {Phys. Rev. B}\ }\textbf {\bibinfo {volume} {111}},\ \bibinfo
  {pages} {054301} (\bibinfo {year} {2025})}\BibitemShut {NoStop}%
\bibitem [{\citenamefont {Tirrito}\ \emph
  {et~al.}(2025{\natexlab{b}})\citenamefont {Tirrito}, \citenamefont
  {Turkeshi},\ and\ \citenamefont {Sierant}}]{tirrito_prl_2025}%
  \BibitemOpen
  \bibfield  {author} {\bibinfo {author} {\bibfnamefont {E.}~\bibnamefont
  {Tirrito}}, \bibinfo {author} {\bibfnamefont {X.}~\bibnamefont {Turkeshi}},\
  and\ \bibinfo {author} {\bibfnamefont {P.}~\bibnamefont {Sierant}},\
  }\bibfield  {title} {\bibinfo {title} {Anticoncentration and
  nonstabilizerness spreading under ergodic quantum dynamics},\ }\href
  {https://doi.org/10.1103/1jzy-sk9r} {\bibfield  {journal} {\bibinfo
  {journal} {Phys. Rev. Lett.}\ }\textbf {\bibinfo {volume} {135}},\ \bibinfo
  {pages} {220401} (\bibinfo {year} {2025}{\natexlab{b}})}\BibitemShut
  {NoStop}%
\bibitem [{\citenamefont {Dowling}\ \emph {et~al.}(2025)\citenamefont
  {Dowling}, \citenamefont {Kos},\ and\ \citenamefont
  {Turkeshi}}]{dowling_prl_2025}%
  \BibitemOpen
  \bibfield  {author} {\bibinfo {author} {\bibfnamefont {N.}~\bibnamefont
  {Dowling}}, \bibinfo {author} {\bibfnamefont {P.}~\bibnamefont {Kos}},\ and\
  \bibinfo {author} {\bibfnamefont {X.}~\bibnamefont {Turkeshi}},\ }\bibfield
  {title} {\bibinfo {title} {Magic resources of the heisenberg picture},\
  }\href {https://doi.org/10.1103/p7xt-s9nz} {\bibfield  {journal} {\bibinfo
  {journal} {Phys. Rev. Lett.}\ }\textbf {\bibinfo {volume} {135}},\ \bibinfo
  {pages} {050401} (\bibinfo {year} {2025})}\BibitemShut {NoStop}%
\bibitem [{\citenamefont {Bejan}\ \emph {et~al.}(2025)\citenamefont {Bejan},
  \citenamefont {Claeys},\ and\ \citenamefont {Yao}}]{bejan2025}%
  \BibitemOpen
  \bibfield  {author} {\bibinfo {author} {\bibfnamefont {M.}~\bibnamefont
  {Bejan}}, \bibinfo {author} {\bibfnamefont {P.~W.}\ \bibnamefont {Claeys}},\
  and\ \bibinfo {author} {\bibfnamefont {J.}~\bibnamefont {Yao}},\ }\href
  {https://arxiv.org/abs/2511.21487} {\bibinfo {title} {Magic spreading under
  unitary clifford dynamics}} (\bibinfo {year} {2025}),\ \Eprint
  {https://arxiv.org/abs/2511.21487} {arXiv:2511.21487 [quant-ph]} \BibitemShut
  {NoStop}%
\bibitem [{\citenamefont {Maity}\ and\ \citenamefont
  {Hamazaki}(2026)}]{maity_prr_2026}%
  \BibitemOpen
  \bibfield  {author} {\bibinfo {author} {\bibfnamefont {S.}~\bibnamefont
  {Maity}}\ and\ \bibinfo {author} {\bibfnamefont {R.}~\bibnamefont
  {Hamazaki}},\ }\bibfield  {title} {\bibinfo {title} {Local spreading of
  stabilizer r\'enyi entropy in a brickwork random clifford circuit},\ }\href
  {https://doi.org/10.1103/68g7-8pdc} {\bibfield  {journal} {\bibinfo
  {journal} {Phys. Rev. Res.}\ }\textbf {\bibinfo {volume} {8}},\ \bibinfo
  {pages} {013324} (\bibinfo {year} {2026})}\BibitemShut {NoStop}%
\bibitem [{\citenamefont {Agrawal}\ \emph {et~al.}(2022)\citenamefont
  {Agrawal}, \citenamefont {Zabalo}, \citenamefont {Chen}, \citenamefont
  {Wilson}, \citenamefont {Potter}, \citenamefont {Pixley}, \citenamefont
  {Gopalakrishnan},\ and\ \citenamefont {Vasseur}}]{agarwal_prx_2022}%
  \BibitemOpen
  \bibfield  {author} {\bibinfo {author} {\bibfnamefont {U.}~\bibnamefont
  {Agrawal}}, \bibinfo {author} {\bibfnamefont {A.}~\bibnamefont {Zabalo}},
  \bibinfo {author} {\bibfnamefont {K.}~\bibnamefont {Chen}}, \bibinfo {author}
  {\bibfnamefont {J.~H.}\ \bibnamefont {Wilson}}, \bibinfo {author}
  {\bibfnamefont {A.~C.}\ \bibnamefont {Potter}}, \bibinfo {author}
  {\bibfnamefont {J.~H.}\ \bibnamefont {Pixley}}, \bibinfo {author}
  {\bibfnamefont {S.}~\bibnamefont {Gopalakrishnan}},\ and\ \bibinfo {author}
  {\bibfnamefont {R.}~\bibnamefont {Vasseur}},\ }\bibfield  {title} {\bibinfo
  {title} {Entanglement and charge-sharpening transitions in u(1) symmetric
  monitored quantum circuits},\ }\href
  {https://doi.org/10.1103/PhysRevX.12.041002} {\bibfield  {journal} {\bibinfo
  {journal} {Phys. Rev. X}\ }\textbf {\bibinfo {volume} {12}},\ \bibinfo
  {pages} {041002} (\bibinfo {year} {2022})}\BibitemShut {NoStop}%
\bibitem [{\citenamefont {Chakraborty}\ \emph {et~al.}(2024)\citenamefont
  {Chakraborty}, \citenamefont {Chen}, \citenamefont {Zabalo}, \citenamefont
  {Wilson},\ and\ \citenamefont {Pixley}}]{chakraborty_prb_2024}%
  \BibitemOpen
  \bibfield  {author} {\bibinfo {author} {\bibfnamefont {A.}~\bibnamefont
  {Chakraborty}}, \bibinfo {author} {\bibfnamefont {K.}~\bibnamefont {Chen}},
  \bibinfo {author} {\bibfnamefont {A.}~\bibnamefont {Zabalo}}, \bibinfo
  {author} {\bibfnamefont {J.~H.}\ \bibnamefont {Wilson}},\ and\ \bibinfo
  {author} {\bibfnamefont {J.~H.}\ \bibnamefont {Pixley}},\ }\bibfield  {title}
  {\bibinfo {title} {Charge and entanglement criticality in a u(1)-symmetric
  hybrid circuit of qubits},\ }\href
  {https://doi.org/10.1103/PhysRevB.110.045135} {\bibfield  {journal} {\bibinfo
   {journal} {Phys. Rev. B}\ }\textbf {\bibinfo {volume} {110}},\ \bibinfo
  {pages} {045135} (\bibinfo {year} {2024})}\BibitemShut {NoStop}%
\bibitem [{\citenamefont {Summer}\ \emph {et~al.}(2026)\citenamefont {Summer},
  \citenamefont {Nico-Katz}, \citenamefont {Dooley},\ and\ \citenamefont
  {Goold}}]{Summer2026}%
  \BibitemOpen
  \bibfield  {author} {\bibinfo {author} {\bibfnamefont {A.}~\bibnamefont
  {Summer}}, \bibinfo {author} {\bibfnamefont {A.}~\bibnamefont {Nico-Katz}},
  \bibinfo {author} {\bibfnamefont {S.}~\bibnamefont {Dooley}},\ and\ \bibinfo
  {author} {\bibfnamefont {J.}~\bibnamefont {Goold}},\ }\bibfield  {title}
  {\bibinfo {title} {Anomalous transport in u(1)-symmetric quantum circuits},\
  }\bibfield  {journal} {\bibinfo  {journal} {npj Quantum Information}\
  }\textbf {\bibinfo {volume} {12}},\ \href
  {https://doi.org/10.1038/s41534-025-01178-8} {10.1038/s41534-025-01178-8}
  (\bibinfo {year} {2026})\BibitemShut {NoStop}%
\bibitem [{\citenamefont {Li}\ \emph {et~al.}(2026{\natexlab{b}})\citenamefont
  {Li}, \citenamefont {Lee}, \citenamefont {Liu}, \citenamefont {Zhang},\ and\
  \citenamefont {Zhong}}]{Li2026_b}%
  \BibitemOpen
  \bibfield  {author} {\bibinfo {author} {\bibfnamefont {H.-Z.}\ \bibnamefont
  {Li}}, \bibinfo {author} {\bibfnamefont {C.~H.}\ \bibnamefont {Lee}},
  \bibinfo {author} {\bibfnamefont {S.}~\bibnamefont {Liu}}, \bibinfo {author}
  {\bibfnamefont {S.-X.}\ \bibnamefont {Zhang}},\ and\ \bibinfo {author}
  {\bibfnamefont {J.-X.}\ \bibnamefont {Zhong}},\ }\bibfield  {title} {\bibinfo
  {title} {Quantum mpemba effect in long-ranged u(1)-symmetric random
  circuits},\ }\href {https://doi.org/10.1103/l4xx-plmv} {\bibfield  {journal}
  {\bibinfo  {journal} {Phys. Rev. B}\ }\textbf {\bibinfo {volume} {113}},\
  \bibinfo {pages} {134310} (\bibinfo {year} {2026}{\natexlab{b}})}\BibitemShut
  {NoStop}%
\bibitem [{\citenamefont {Bera}\ and\ \citenamefont
  {Singha~Roy}(2020)}]{bera_pra_2020}%
  \BibitemOpen
  \bibfield  {author} {\bibinfo {author} {\bibfnamefont {A.}~\bibnamefont
  {Bera}}\ and\ \bibinfo {author} {\bibfnamefont {S.}~\bibnamefont
  {Singha~Roy}},\ }\bibfield  {title} {\bibinfo {title} {Growth of genuine
  multipartite entanglement in random unitary circuits},\ }\href
  {https://doi.org/10.1103/PhysRevA.102.062431} {\bibfield  {journal} {\bibinfo
   {journal} {Phys. Rev. A}\ }\textbf {\bibinfo {volume} {102}},\ \bibinfo
  {pages} {062431} (\bibinfo {year} {2020})}\BibitemShut {NoStop}%
\bibitem [{\citenamefont {Pusz}\ and\ \citenamefont
  {Woronowicz}(1978)}]{Pusz78}%
  \BibitemOpen
  \bibfield  {author} {\bibinfo {author} {\bibfnamefont {W.}~\bibnamefont
  {Pusz}}\ and\ \bibinfo {author} {\bibfnamefont {S.~L.}\ \bibnamefont
  {Woronowicz}},\ }\bibfield  {title} {\bibinfo {title} {Passive states and
  {KMS} states for general quantum systems},\ }\href@noop {} {\bibfield
  {journal} {\bibinfo  {journal} {Commun. Math. Phys.}\ }\textbf {\bibinfo
  {volume} {58}} (\bibinfo {year} {1978})}\BibitemShut {NoStop}%
\bibitem [{\citenamefont {Viscardi}\ \emph {et~al.}(2026)\citenamefont
  {Viscardi}, \citenamefont {Dalmonte}, \citenamefont {Hamma},\ and\
  \citenamefont {Tirrito}}]{Viscardi26}%
  \BibitemOpen
  \bibfield  {author} {\bibinfo {author} {\bibfnamefont {M.}~\bibnamefont
  {Viscardi}}, \bibinfo {author} {\bibfnamefont {M.}~\bibnamefont {Dalmonte}},
  \bibinfo {author} {\bibfnamefont {A.}~\bibnamefont {Hamma}},\ and\ \bibinfo
  {author} {\bibfnamefont {E.}~\bibnamefont {Tirrito}},\ }\bibfield  {title}
  {\bibinfo {title} {Interplay of entanglement structures and stabilizer
  entropy in spin models},\ }\href
  {https://doi.org/10.21468/SciPostPhysCore.9.1.012} {\bibfield  {journal}
  {\bibinfo  {journal} {SciPost Phys. Core}\ }\textbf {\bibinfo {volume} {9}},\
  \bibinfo {pages} {012} (\bibinfo {year} {2026})}\BibitemShut {NoStop}%
\end{thebibliography}%
\end{document}